\documentclass[preprint,12pt,eqsecnum]{aastex}           
\usepackage{natbib,graphics}

\newcommand{\be}{\begin{equation}}
\newcommand{\ee}{\end{equation}}

\newcommand{\bea}{\begin{eqnarray}}
\newcommand{\eea}{\end{eqnarray}}

\newcommand{\alphav}{\alpha_*}

\begin{document}
\title{Relativistic Diskoseismology. III. Low-Frequency Fundamental P--modes} 

\author{Manuel Ortega-Rodr\'{\i}guez\altaffilmark{1}}
\affil{Department of Applied Physics and Gravity Probe B, \\
Stanford University, Stanford, CA 94305--4090}
\author{Alexander S. Silbergleit\altaffilmark{2}}
\affil{Gravity Probe B, 
W. W. Hansen Experimental Physics Lab, \\
Stanford University, Stanford, CA 94305--4085}
\and\author{Robert V. Wagoner\altaffilmark{3}} 
\affil{Department of Physics and Center for Space Science and Astrophysics \\ 
Stanford University, Stanford, CA 94305--4060}

\altaffiltext{1}{mortega@cariari.ucr.ac.cr}
\altaffiltext{2}{gleit@stanford.edu} 
\altaffiltext{3}{wagoner@stanford.edu}

\begin{abstract}

We extend our investigation of the normal modes of small adiabatic oscillations of relativistic barotropic thin accretion disks to the inertial--pressure ($p$) modes. We focus here on the lowest frequency fundamental $p$--modes, those with no axial or vertical nodes in their distribution. Through a variety of analyses, we obtain closed-form expressions for the eigenfrequencies and eigenfunctions. These depend upon the luminosity and viscosity parameter of the disk, as well as the mass and angular momentum of the black hole, via detailed formulae for the speed of sound. The effect of a torque on the inner edge of the disk is also included. 
We compare the $p$--mode properties to those of the $g$-- and $c$--modes.

\end{abstract}


\section{Introduction}

All perturbations of equilibrium models of accretion disks can be described in terms of their normal modes of oscillation, which our group has been studying during the last decade, following the pioneering work of \citet{kf} [see \citet{kato} for a review]. 
Approaches involving instead more local (e.g., dispersion relation) considerations have been extensively used to study the propagation of waves in disks , with the inclusion of buoyancy, atmospheres, and magnetic fields [see, e.g., \citet{ol} and references therein]. Their classification of modes is similar to those for stars, and different than ours. Our focus is rather on global modes, and in the determination of eigenfrequencies and their applications. All the modes we study have discrete, real eigenfrequencies. Viscous damping (or growth) is assumed to introduce a relatively small imaginary contribution.

Effects of general relativity trap most of these modes near the inner edge of accretion disks around compact objects. The strong gravitational fields that are required can be produced by a black hole or by neutron stars that are sufficiently compact (with a soft equation of state) and weakly magnetized to produce a gap between the surface of the star and the innermost stable orbit of the accretion disk. Although we shall not explicitly consider such neutron stars here, the results obtained will also apply to them to first order in the dimensionless angular momentum parameter $a=cJ/GM^2$, since their exterior metric is identical to that of a black hole to that order. 

The subject of `relativistic diskoseismology' has been reviewed by \citet{k98} and \citet{w}. 
In this paper we shall focus on the fundamental low frequency (LF) pressure-driven p--modes: those axisymmetric modes with no nodes in their vertical distribution and with the smallest possible eigenfrequency. We shall briefly compare these modes with the `corrugation'(c) modes \citep{swo} and the `gravity'(g) modes \citep{per}.  An investigation of the other types of p--modes, including those which can cover the entire disk, will be published separately \citep{osw}.
The study of p--modes will complete the last phase of our analysis of the normal modes of oscillation of `standard' relativistic thin accretion disks.

\section{Basic Assumptions and Equations}

\subsection{Structure of the unperturbed accretion disk}

We take $c=1$, and express all distances in units of $GM/c^2$ and all frequencies in units of $c^3/GM$ (where $M$ is the mass of the central body) unless otherwise indicated. We employ the Kerr metric to study a thin accretion disk, neglecting its self-gravity. The stationary ($\partial/\partial t=0$), symmetric about the midplane $z=0$, and axially symmetric ($\partial/\partial\varphi$=0) equilibrium disk is taken to be described by the standard relativistic thin 
disk model \citep{nt,pt}. The velocity components $v^r=v^z=0$, and the disk semi-thickness $h(r)\sim c_s/\Omega\ll r$, where $c_s(r,z)$ is the speed of sound. The key frequencies, associated with free-particle orbits, are 
\bea \label{3freq}
\Omega(r) & = & (r^{3/2}+a)^{-1}\; , \nonumber \\
\Omega_\perp(r) & = & \Omega(r)\left(1-4a/r^{3/2}+3a^2/r^2\right)^{1/2}\; , \nonumber \\
\kappa(r) & = & \Omega(r)\left(1-6/r+8a/r^{3/2}-3a^2/r^2\right)^{1/2}\; ;\label{eq:1}
\eea
the rotational, vertical epicyclic, and radial epicyclic frequencies, respectively. The angular momentum parameter $a$ is less than unity in absolute value. 

The inner edge of the disk is at approximately the radius of the last stable free-particle circular orbit $r=r_i(a)$, where the epicyclic frequency $\kappa(r_i)=0$. This radius $r_i(a)$ is a decreasing function of $a$, from $r_i(-1)=9$ through $r_i(0)=6$ to $r_i(1)=1$. So all the relations we use are for $r>r_i$, where $\kappa(r)>0$. Note, in particular, that
\be
\Omega(r)>\Omega_\perp(r)>\kappa(r) \; ,\quad a>0 \; ;\qquad \Omega_\perp(r)>\Omega(r)>\kappa(r) \; ,\quad a<0 \; .
\label{eq:2}
\ee
For $a=0$ (a non-rotating black hole), $\Omega(r)=\Omega_\perp(r)>\kappa(r)$. We will also employ the outer disk radius, $r_{o}$. 
For stellar mass black holes in low-mass X-ray binaries, orbital separations require that $r_{o}\lesssim 10^5 M_\sun/M$.
The properties of the outer disk have not been greatly constrained from observations, but the matter impacting from the companion star is likely to significantly disturb the outer region of the disk. We neglect such complications to our equilibrium model in this exploratory treatment of the outer modes. 

To simplify the analysis, we here consider barotropic disks [$p=p(\rho)$, vanishing buoyancy frequency; a generalization to a small non-zero buoyancy may be found in \citet{swo}, section 4.3, and \citet{per}]. In this case hydrostatic equilibrium provides the vertical density and pressure profiles
\be
\rho=\rho_0(r)(1-y^2)^{g}\; ,\quad p=p_0(r)(1-y^2)^{g+1}\;, \quad 
g\equiv1/(\Gamma-1) >0 \; , \label{eq:3}
\ee
where $\Gamma>1$ is the adiabatic index (for brevity and convenience, we will use parameters $g$ and $\Gamma$ alternatively). One has $\Gamma=4/3$ within any radiation pressure dominated region of the disk, and $\Gamma=5/3$ within any gas pressure dominated region.The disk surfaces are at $y=\pm1$, with $y$ related to the vertical coordinate $z$ by
\[ 
y={z\over h(r)}\,\sqrt{\Gamma-1\over2\Gamma} \; ,
\]
and $h(r)$ specified by equation (2.19) of \citet{per}. 
More information on the unperturbed disk is given below.

\subsection{Equations for the disk perturbations}

To investigate the eigenmodes of the disk oscillations, we apply the general relativistic formalism that \citet{il} developed for perturbations of purely rotating perfect fluids.  Neglecting the self-gravity of the disk is usually a very good approximation, as will be discussed in section 8. 
Viscosity corrections to the perturbations are of order $\alphav$ (the usual viscosity parameter, not to be confused with the function $\alpha$ introduced below) and are assumed small. (However, they are fully included in the equilibrium model via $\alphav$.)
One can then express the Eulerian perturbations of all physical quantities through a single function $\delta V\propto\delta p/\rho$ which satisfies a second-order partial differential equation. Due to the stationary and axisymmetric background, the angular and time dependences are factored out as $\delta V = V(r,z)\exp[i(m\phi + \sigma t)]$, where $\sigma$ is the eigenfrequency. Then the assumption of strong variation of modes in the radial direction (characteristic radial wavelength $\lambda_r \ll r$) ensures the approximate WKB separability of variables in the partial differential equation for the functional amplitude $V(r,z) = V_r(r)V_y(r,y)$. The function $V_y$ varies slowly with $r$. 

The resulting ordinary differential equations for the vertical ($V_y$) and radial ($V_r$) eigenfunctions
are [see \citet{nw92}, \citet{per} and \citet{swo} for details]:  
\bea
(1-y^2)\,\frac{d^2V_y}{d y^2} - {2gy}\,\frac{d V_y}{d y} + {2g\omega_*^2}\,
\left[1 -\left(1-\frac{\Psi}{\omega_*^2}\right)\left(1-y^2\right)\right]V_y = 0 \; , 
\label{eq:4} \\
\frac{d^2 V_r}{dr^2} - \frac{1}{(\omega^2-\kappa^2)} \left[\frac{d}{dr}
(\omega^2-\kappa^2)\right]\frac{dV_r}{dr} +\alpha^2(\omega^2-\kappa^2)\left(1 -
\frac{\Psi}{\omega_*^2}\right)V_r = 0 \; . \label{eq:5}
\eea 
The coefficient $\alpha(r)$ is defined as
\be \label{defalpha}
\alpha(r) \equiv \beta_{rel} \sqrt{g_{rr}}/c_s(r,0) \; , \label{alpha}
\ee
where $\beta_{rel} = dt/d\tau$ and $g_{rr}$ is a Kerr metric coefficient in Boyer-Lindquist coordinates.

As in \citet{swo},  we assume that
\be
\alpha(r)=\gamma(r)\frac{r^{\mu+\nu}}{(r-r_i)^{\mu}} \; , \qquad 
0 \leq \mu < 1/2 \; , \quad \nu\geq 0 \; , \label{3.7}
\ee
where $\gamma(r)$ is some function bounded from above and away from zero, varying slowly with radius, and tending at infinity to a limit $\gamma(\infty) \equiv \gamma_\infty>0$. Thus for very large radii $\alpha(r)$ is effectively a nonnegative power,
\be
\alpha(r)\approx\gamma_\infty r^{\nu}\; , \label{outer}
\ee
while near the inner edge of the disk 
\be
\alpha(r)\approx{\alpha_i}{(r-r_i)^{-\mu}}\; ,\qquad \alpha_i\equiv\gamma(r_i)r_i^{\mu+\nu}\;. \label{inner}
\ee
The singularity at $r=r_i$ is due to the fact that the pressure at the inner edge vanishes in a typical disk model [$p\propto (r-r_i)^k$] if the torque does, in which case $\mu=(\Gamma-1)k/2\Gamma$. Typically, $k=2$ and $\Gamma=5/3$, giving $\mu=2/5$. If even a small torque is applied to the disk at the inner edge, which is likely due to magnetic stress \citep{hk}, the pressure and the speed of sound become nonzero at $r=r_i$. This corresponds to a nonsingular $\alpha(r)$ with $\mu=0$. 

It is also true that  $\nu<1$ in most disk models; if it is less than one--half, some limitation on the disk size (given $\gamma_\infty$, and vice versa) applies. Indeed, as stated in section 2.1, 
\[
h(r)\sim c_s/\Omega\sim 1/\alpha\Omega\ll r \; .
\]
This relation, according to formula (\ref{outer}) for a large outer radius $r_o\gg r_i$, converts into
\be
r_o^{1/2-\nu}\ll\gamma_\infty \; ,
\label{3.10}
\ee
and we have the following three cases:

1. For $\nu>1/2$, the outer radius of the disk should satisfy $r_o\gg\gamma_\infty^{-1/(\nu-0.5)}$.

2. For $\nu=1/2$, it is not the size of the disk, but the speed of sound that is limited by the requirement $\gamma_\infty\gg1$.

3. For $0\leq\nu<1/2$ the outer radius is limited by $r_o\ll\gamma_\infty^{1/(0.5-\nu)}$.

\noindent One should specifically bear in mind this last limitation in the following analysis, since $0\leq\nu<1/2$ in most accretion disk models.

Together with the appropriate homogeneous boundary conditions [discussed by \citet{per}, \citet{swo}, and below in sections 3, 4], equations (\ref{eq:4}) and (\ref{eq:5}) generate the vertical and radial eigenvalue problems, respectively.  The radial boundary  conditions depend on the type of mode and its capture zone, as discussed in the following sections. The coefficient $\alpha$, the vertical eigenfunction $V_y$ and  eigenvalue (separation function) $\Psi$ vary slowly with radius (comparable to that of the equilibrium model),
as does the (dimensionless) ratio of the corotation frequency $\omega(r)$ and $\Omega_\perp(r)$: 
\be
\omega_*(r)\equiv\omega(r)/\Omega_\perp(r) \; , \qquad\omega(r)\equiv\sigma+m\Omega(r) \; . \label{eq:6}
\ee
(Note that $\Psi$ and $\omega_*$ depend also on the eigenfrequency $\sigma$.)

Our goal is to determine the vertical eigenvalues $\Psi(r,\sigma)$ and the corresponding spectrum of eigenfrequencies $\sigma$ from the two eigenvalue problems for equations (\ref{eq:4}) and (\ref{eq:5}). Along with the angular mode number $m$, we employ $j$ and $n$ for the vertical and radial mode numbers (number of nodes in the corresponding eigenfunction), respectively, as in \citet{per} and \citet{swo}.  
 
\subsection{Classification of Modes}

One can see from the radial equation (\ref{eq:5}) that a mode oscillates in a domain of $r>r_i$ where the quantity $(\omega^2-\kappa^2)(1 -\Psi/\omega_*^2)$ is positive. The first factor here (for $\sigma$ in the allowed range) changes sign twice at the points $r=r_{\pm}(m,a,\sigma),\,\,r_i<r_-\leq r_+ $, the Lindblad resonances determined by equations (\ref{3freq}) and (\ref{eq:6}). This factor is negative between $r_-$ and $r_+$ and positive otherwise [see fig. 1 in \citet{swo}].  Thus the classification of modes is based on the behavior of the second factor, containing the eigenvalue. 

There are essentially three possibilities, corresponding to the following types of modes: 

\noindent$\bullet$ {\it g--modes}: $\Psi>$(typically\ $\gg )\,\omega_*^2$; capture zone $r_-<r<r_+$. 

\noindent$\bullet$ {\it p--modes}: $\Psi<$(typically\ $\ll )\,\omega_*^2$; capture zone $r_i<r<r_-$ or $r_+<r<r_o$.

\noindent$\bullet$ {\it c--modes}: $\Psi = \omega_*^2$, for at least one value of the radius. 

\noindent The capture, or trapping, zone is where the mode oscillations occur.

The g (inertial-gravity)--modes are centered on the radius $r_m$ where $r_-=r_+$, corresponding to the maximum eigenfrequency $|\sigma|$. The corotating eigenfrequencies $|\omega|$ corresponding to the lowest mode numbers are close to $\kappa(r_m)$ [see \citet{per} for details]. 

The c(corrugation)--modes which have so far been investigated are of a special sort: $\Psi-\omega_*^2\approx 0$ throughout their whole capture domain. They are non-radial ($m\not=0$, typically $m=\pm1$) vertically incompressible waves near the inner edge of the corotating disk that precess around the angular momentum of the black hole. Their fundamental frequency is shown to coincide with the Lense-Thirring \citep{lt} frequency (produced by the dragging of inertial frames generated by the angular momentum of the black hole) at their outer capture zone boundary $r_c(m,a)$ in the appropriate slow-rotation limit \citep{swo}. 

In this paper we investigate p (inertial-pressure)--modes, which have been studied since the pioneering analysis of \citet {kf}. However, the only previous relativistic treatment was by  \citet{p}. As mentioned in the introduction, in this paper we concentrate on the properties of the fundamental ($m=j=0$) low frequency p--modes.

\section{Perturbative Solution of the Vertical Eigenvalue Problem}

We will use the self-adjoint form of the vertical equation (\ref{eq:4}),
\be
\frac{d}{dy}
\left[
\left(1-y^2\right)^{g}\frac{dV_y}{dy}
\right] +
2g\omega_*^2\,
\left[
1 - 
\left(1-\frac{\Psi}{\omega_*^2}\right)\left(1-y^2\right)
\right]
\left(1-y^2\right)^{g-1}V_y = 0 \; . \label{4.1}
\ee
For the lowest angular mode number $m=0$ we have the corotation frequency coinciding with the eigenfrequency, $\omega^2=\sigma^2$ [from equation (\ref{eq:6})]. Hence we can rewrite the above equation as
\be
{\cal L}V_y=
-\left(\sigma^2/\Omega_\perp^2\right)2g\,\left[
1 - \left(1-{\Psi}/{\omega_*^2}\right)\left(1-y^2\right)
\right]
\left(1-y^2\right)^{g-1}V_y \;. \label{4.2}
\ee
Here the Hermitian operator
\be
{\bf\cal L}\,\cdot\equiv
\frac{d}{dy}
\left[
\left(1-y^2\right)^{g}\frac{d}{dy}\right]\,\cdot
\; \label{4.3}
\ee
acts on those smooth enough functions of $y$ on the interval $(-1,\,1)$ which are bounded together with their derivative at $y=\pm1$. The last requirement is dictated by physics \citep{per}. It also turns out to be a natural boundary condition for the singular ODE (\ref{4.1}).

We are interested in this first paper in the low-frequency (LF) p--modes. We specify the LF range by requiring that the following frequency ratio be small enough throughout the appropriate capture zone:
\be
\sigma^2/\Omega_\perp^2(r,a) \lesssim 1\; .\label{4.4}
\ee
The frequency $\Omega_\perp(r,a)$ decreases monotonically with radius for the rotation parameter $a$ within the interval 
$-1\leq a\leq a_\perp\approx 0.953$; for  $a>a_\perp$ it has a single maximum near the inner edge, but is still decreasing far out as $r^{-3/2}$. Thus for the outer LF p--modes captured within $r_+<r<r_o$, the requirement (\ref{4.4}) needs to be checked only at the outer edge of the disk ($r=r_o$). The LF inner fundamental p--modes are concentrated very close to the inner edge, so the requirement (\ref{4.4}) needs to be checked only at $r=r_i$ for them. 

Note that from equation (\ref{eq:1}),
\[
\Omega_\perp(r_i,a)\bigl|_{a=1}=\Omega_\perp(1,1)=0\; .
\]
Hence for very rapidly rotating black holes ($a>a_\perp$) the inner modes we are studying are not well defined. Some more specific constraints on the parameters for both the outer and inner LF p--modes are given later in this and following sections.

Having set up the basic small parameter (\ref{4.4}), we solve the vertical eigenvalue problem by means of perturbations,
\be
1-\frac{\Psi}{\omega_*^2}=\chi_0+\left(\frac{\sigma}{\Omega_\perp}\right)^2\chi_1+
\left(\frac{\sigma}{\Omega_\perp}\right)^4\chi_2+\dots\; ;\label{4.5}
\ee
\be
 V_y=V^{(0)}+\left(\frac{\sigma}{\Omega_\perp}\right)^2V^{(1)}+
\left(\frac{\sigma}{\Omega_\perp}\right)^4V^{(2)}+\dots,\label{4.6}
\ee
with the coefficients $\chi_k$ and functions $V^{(k)}(y),\,\,k=0,1,2,\dots,$ to be determined. (Recall that both quantities are, in general, slowly varying functions of the radius, as is $\Omega_\perp$.) The substitution of these expansions in equation (\ref{4.2}) leads to the following chain of equations
\be
{\cal L}V^{(0)}=0 \; , \label{4.7}
\ee
\be
{\cal L}V^{(k)}=-f_k \; ; \label{4.8}
\ee
\[
f_k(y)\equiv 2g\,\left[\left(1-y^2\right)^{g-1}V^{(k-1)}(y)-
\left(1-y^2\right)^{g}\sum_{l=0}^{k-1}\,\chi_lV^{(k-l-1)}(y)\right]\; ,
\quad k\geq1\; ,
 \]
with the boundary conditions that all $V^{(k)}(y),\,\,k\geq0,$ and their first derivatives be finite at $y^2=1$. Note that the right hand side $f_k(y)$ is a linear combination of $V^{(i)}(y)$ whose coefficients depend on $\chi_{i}$, with $0\leq i\leq k-1$.

The only solution to the described `ground state' $(\sigma = 0)$ problem is
\be
V^{(0)}(y)= {\rm constant} \equiv 1\; .
 \label{4.9}
\ee
This shows that, at least to lowest order, the vertical eigenfunction has no nodes ($j=0$) and is even. A straightforward  symmetry and induction argument immediately demonstrates that all the corrections $V^{(k)}(y)$, and hence the vertical eigenfunction $V_y(y)$, are also even. This allows us to work on the  halved interval $y\in(0,\,1)$ with the proper symmetry condition at $y=0$. It is convenient to introduce the notation $\langle f,g \rangle\equiv\int\limits_0^1f(y)g(y)dy$ for the scalar product.

The procedure of successive solutions of equations (\ref{4.8}) is standard. For a given $k\geq1$, provided that all the previous equations have been solved and the eigenvalue corrections $\chi_0,\,\chi_1,\,\dots,\chi_{k-2}$ already found, we first determine $\chi_{k-1}$ from the solvability criterion (the operator ${\cal L}$ is Hermitian),
\be
\langle V^{(0)},f_k\rangle=-\langle V^{(0)},{\cal L}V^{(k)}\rangle=-\langle {\cal L}V^{(0)},V^{(k)}\rangle=0\; .
 \label{4.10}
\ee
We then find $V^{(k)}(y)$ by direct integration of equation (\ref{4.8}), with a now completely determined right hand side. Of course, $V^{(k)}(y)$ is originally defined up to an additive term proportional to $V^{(0)}(y)={\rm const}$; it is usually made unique by means of the orthogonality condition
\be
\langle V^{(0)},V^{(k)}\rangle=0\; ,\qquad k=1,2,\dots\; .
 \label{4.11}
\ee

We begin by solving the problem for $k=1$. According to expression (\ref{4.9}), the r.h.s. of equation (\ref{4.8}) for the first correction is
\[
f_1(y)= 2g\,
\left[
1 - \chi_0\left(1-y^2\right)
\right]
\left(1-y^2\right)^{g-1}\; ,
\]
so condition (\ref{4.10}) in this case becomes
\be 
0=\langle V^{(0)},f_1\rangle=
 2g\,\int_0^1
\left[
1 - \chi_0\left(1-y^2\right)
\right]
\left(1-y^2\right)^{g-1}dy\; , 
 \label{4.12}
\ee
which allows us to easily calculate
\be
\chi_{0}=1+1/(2g)=(\Gamma+1)/2\; .
 \label{4.13}
\ee
>From the definition (\ref{4.3}) of ${\cal L}$, equation (\ref{4.8}) for $V^{(1)}(y)$ after one integration reduces to
\[
\frac{dV^{(1)}}{dy}=2g\,
\left(1-y^2\right)^{-g}\int_y^1\left[
1 - \chi_0\left(1-\eta^2\right)
\right]
\left(1-\eta^2\right)^{g-1}d\eta\; ,
\]
which is finite at $y=1$. Also, due to equation (\ref{4.12}), the derivative goes to zero at $y=0$, as it should by symmetry. It turns out that the 
last equation simplifies remarkably to just ${dV^{(1)}}/{dy}=y$, and the first correction to the eigenfunction satisfying condition (\ref{4.11}) is thus
\be
V^{(1)}(y)=0.5\left(y^2-1/3\right)\; . \label{4.14}
\ee

To see how the expansion (\ref{4.5}) is generated, we now want to calculate $\chi_1$; that is, to explore the solvability criterion (\ref{4.10}) for $k=2$. According to equations (\ref{4.8}) and (\ref{4.14}), the r.h.s. of the equation for $V^{(2)}(y)$ becomes
\[
f_2(y)= g\,
\left[
\left(y^2-1/3\right)
\left[
1 - \chi_0\left(1-y^2\right)
\right]
\left(1-y^2\right)^{g-1}-
2\left(1-y^2\right)^{g}\chi_1
\right] \; .
 \]
Using this and formula (\ref{4.13}), from equation (\ref{4.10}) with $k=2$ it is not very difficult to find that
\be
\chi_{1}=\frac{1}{2g\left(2g+3\right)}=
\frac{\left(\Gamma-1\right)^2}{2\left(3\Gamma-1\right)}\; .
 \label{4.15}
\ee
So the equalities (\ref{4.5}), (\ref{4.6}) and (\ref{4.13})---(\ref{4.15}) provide together the two term expansions for both the vertical eigenvalue and eigenfunction of the LF fundamental p--modes, namely:
\be
1-\frac{\Psi}{\omega_*^2}=\frac{\Gamma+1}{2}+
\left(\frac{\sigma}{\Omega_\perp}\right)^2\frac{\left(\Gamma-1\right)^2}{2\left(3\Gamma-1\right)}+\dots\;\label{4.16}\; ,
\ee
\be
 V_y(y)=1+\left(\frac{\sigma}{\Omega_\perp}\right)^2\frac{1}{2}\left(y^2-\frac{1}{3}\right)+\dots \label{4.17}\; .
\ee
Apart from everything else, the first of these expressions gives a plausible criterion of how small our expansion parameter (\ref{4.4}) should be. For our applications it is sufficient to have the second term in the expansion (\ref{4.16}) smaller than the leading one. That is,
\be
\left(\frac{\sigma}{\Omega_\perp}\right)^2\ll \frac{\chi_0 }{\chi_1 }=
\frac{\left(\Gamma+1\right)\left(3\Gamma-1\right)}{\left(\Gamma-1\right)^2};\qquad
\frac{\chi_0 }{\chi_1 }>15,\quad 1<\Gamma<2\; . \label{4.18}
\ee
As mentioned above, this should be verified for $r=r_o$ in the case of outer modes and for $r=r_i$ in the case of inner ones.

So, to lowest order
\be
1-\frac{\Psi}{\omega_*^2}\approx\frac{\Gamma+1}{2}\;\label{4.19}
\ee
is independent of the radius
(if $\Gamma$ is). The value of this constant for $\Gamma=4/3$ agrees with the previous numerical results of \citet{p}, which we have checked by our own numerical integration. Note also that the vertical eigenvalue $\Psi$ defined by equation (\ref{4.16}) is negative,
\[
\frac{\Psi}{\omega_*^2}\approx-\left(\frac{\Gamma-1}{2}\right)-
\left(\frac{\sigma}{\Omega_\perp}\right)^2\frac{\left(\Gamma-1\right)^2}{2\left(3\Gamma-1\right)}<0 \; ; 
\]
and varies slowly with radius, since $\omega_*(r)$ and $\Omega_\perp(r)$ do so.

\section{Transformation of the Radial Equation and Boundary Conditions}

It turns out that the radial equation (\ref{eq:5}) is not convenient for analytical solution, be it a WKB approach as in \citet{per} and \citet{swo}, or some other asymptotic methods used in the sequel. To convert it, we first recall its self-adjoint form, namely
\be
\frac{d}{dr}
\left[
\frac{1}{(\omega^2-\kappa^2)}\frac{dV_r}{dr}
\right] +
\alpha^2\left(1 - \frac{\Psi}{\omega_*^2}\right)V_r = 0 \; ; \label{3.1}
\ee
and then [as in \citet{per}, section 3.1] invoke a definition of a new eigenfunction
\be
W=\frac{1}{(\omega^2-\kappa^2)}\frac{dV_r}{dr}\; . \label{3.2}
\ee
In terms of $W(r)$, equation (\ref{3.1}) becomes
\be
V_r=-\frac{1}{\alpha^2\left(1 - \Psi/\omega_*^2\right)}\frac{dW}{dr} \; ; \label{3.3}
\ee
and by substituting equation (\ref{3.3}) into the definition (\ref{3.2}) we obtain the equation 
\be
\frac{d}{dr}
\left[\frac{1}{\alpha^2\left(1 - \Psi/\omega_*^2\right)}
\frac{dW}{dr}\right]+(\omega^2-\kappa^2)W=0 \; \label{3.4}
\ee
for the new radial function. This is the exact equation for $W(r)$, equivalent to equation (\ref{3.1}) for $V_r(r)$. In contrast, an approximate equation for $W(r)$ was used in \citet{per} [see formula (3-3) there]. It was obtained by neglecting the derivative of $\alpha^2\left(1 - \Psi/\omega_*^2\right)$. Fortunately, both equations lead to the same result for g--modes to lowest WKB order.

In the present case of p--modes we introduce the radial independent variable 
\be
\tau=\int_{r_*}^r\alpha^2(r^{'})
\left[1 - \frac{\Psi(r^{'})}{\omega_*^2(r^{'})}\right]
\,dr^{'}\;, \qquad
r_{*}={\rm const}\; .\label{3.5}
\ee
 Since by definition of the p--mode $1 -\Psi/\omega_*^2>0$ in its capture zone, $\tau(r)$ is a monotonic function of the radius there, so the inverse map from $\tau$ to $r$ is well defined. In terms of $\tau$, equation (\ref{3.4}) reduces immediately to the standard WKB form that we need:
\be
\frac{d^2W}{d\tau^2}+S(\tau)W=0\; , \qquad
S(\tau)\equiv\frac{\omega^2-\kappa^2}
{\alpha^2\left(1 - \Psi/\omega_*^2\right)}  \; .
\label{3.6}
\ee
[This particular reduction of a general second order ODE to the WKB form was suggested by \citet{pon}.]
The coefficient $S(\tau)$ is positive in the inner, outer, and full--disk p--mode capture zones, $r_i<r<r_-$, $r_+<r<r_o$, and  $r_i<r<r_o$, respectively. The boundaries $r_{\pm}$ are the turning points of equation (\ref{3.6}). Expressed via the $\tau$ variable, the corresponding capture zones are $\tau_i<\tau<\tau_-$, $\tau_+<\tau<\tau_o$, and $\tau_i<\tau<\tau_o$. We use the natural notation $\tau_\lambda=\tau(r_\lambda)$ for $\lambda=\pm,\,i,\,o$. The function $S(\tau)$ is completely specified, because $\alpha(r)\propto c_s^{-1}(r,0)$, $\omega(r)$, $\omega_*(r)$ and $\kappa(r)$ are given by equations (\ref{3.7}),
(\ref{eq:6}), and (\ref{eq:1}), and $r=r(\tau)$ is the inverse of the function (\ref{3.5}). 

We now move on to the radial boundary conditions. The physical processes and conditions at the edge of the disk are uncertain, but we assume that there is some reflectivity of the mode there so that a boundary condition is appropriate. We parameterize (as usual) the lack of knowledge of this boundary condition by setting the most general conditions there:
\[
\cos \theta_i\; \frac{dV_r}{dr} - \sin\theta_i \; V_r \Biggl|_{r = r_i} = 0\;,\qquad
\cos \theta_o\; \frac{dV_r}{dr} - \sin\theta_o \; V_r \Biggl|_{r = r_o} = 0\; ,
\]
with $-\pi/2<\theta_{i,o}\leq\pi/2$. Because of equations (\ref{3.2}), (\ref{3.3}) and (\ref{3.5}), this translates into the following boundary conditions in terms of $\tau$ and $W(\tau)$:
\be
\omega^2(r_i)\cos \theta_i \; W 
+  \sin \theta_i \; \frac{dW}{d\tau} \Biggl|_{\tau = \tau_i} = 0\; , \label{3.11}
\ee
\be
\left[\omega^2(r_o)-\kappa^2(r_o)\right]\cos \theta_o \; W 
+  \sin \theta_o \; \frac{dW}{d\tau} \Biggl|_{\tau = \tau_o} = 0\; , \label{3.12}
\ee
recalling that $\kappa(r_i)=0$. 

This is the complete set of boundary conditions for the radial equation (\ref{3.6}) for  the full--disk modes. However, for the inner and outer p--modes only one of these conditions is employed, since the other boundary is formed by one of the turning points, either $\tau_-$ or $\tau_+$. In both cases we require the solution to decay outside the turning point boundary of the trapping zone; that is, for $\tau=\tau_-+0$ in the inner and for $\tau=\tau_+-0$ in the outer case. This requirement concludes the formulation of the eigenvalue problem for $W(\tau)$ for p--modes of all types.

\section{Outer Low-Frequency Fundamental P--Modes}

Since $m=0$ and we assume a small eigenfrequency, $r_+\approx\sigma^{-2/3}$. This turning point should be within the disk, so
\be\label{4.20}
r_o>r_+\left(\approx\sigma^{-2/3}\right)\gg r_i>1\; .
\ee
Thus, from equations (\ref{eq:1}) and (\ref{outer}), for all points in the capture zone $r_+\leq r \leq r_o$ we have (to lowest order)
\be\label{4.21}
\kappa(r)\approx\Omega_\perp(r)\approx\Omega(r)\approx r^{-3/2}\;,\qquad
\alpha(r)\approx\gamma_\infty r^{\nu}\;.
\ee

The inequalities (\ref{4.20}) provide the eigenfrequency range
\[
r_o^{-3/2}< \sigma \ll r_i^{-3/2}\;.
\]
However, the validity of condition (\ref{4.18}) for the vertical eigenvalue approximation (\ref{4.19}) at the outer edge gives, in view of equation (\ref{4.21}), a much more restrictive upper bound on the eigenfrequency:
\be\label{4.22}
\frac{1}{r_o^{3/2}}< \sigma \ll 
\frac{\sqrt{\left(\Gamma+1\right)\left(3\Gamma-1\right)}}{\left(\Gamma-1\right)}\,\frac{1}{r_o^{3/2}}
\ee
For example, if $\Gamma=5/3$ this reduces to
\[
r_o^{-3/2}< \sigma \ll 5r_o^{-3/2}\;.
\]
Therefore the only situation which makes our assumptions consistent is 
\be\label{4.23}
\sigma=r_o^{-3/2}(1+\epsilon)\; ,\qquad \quad 
r_+= r_o\left[1-2\epsilon/3+{\cal O}(\epsilon^2)\right]\; , 
\ee
with some small enough positive $\epsilon$ to be found. Note that the upper bound on the eigenfrequency in expression (\ref{4.22}) stems from the use of the simplest approximation (\ref{4.18}), which nevertheless is valid if the eigenfrequency of the mode is low enough. Hence the whole frequency range (\ref{4.22}) is an immediate implication of the requirement that the eigenfrequency of the outer fundamental p--mode is low.

For the outer p--modes the radial eigenvalue problem is considered on the interval $r_+<r<r_o$. Hence it is convenient to chose $r_*=r_+$ in equation (\ref{3.5}), so that
\be\label{4.24}
\tau=\int_{r_+}^r\alpha^2(r^{'})
\left[1 - \frac{\Psi(r^{'})}{\omega_*^2(r^{'})}\right]
\,dr^{'}\;, \quad
\tau_{+}=0\;,\quad
\tau_o=\int_{r_+}^{r_o}\alpha^2(r^{'})
\left[1 - \frac{\Psi(r^{'})}{\omega_*^2(r^{'})}\right]
\,dr^{'} \;, 
\ee
and the capture zone is $0<\tau<\tau_o$. In our case ($m=0$, low frequency) we replace the vertical eigenvalue with its expression (\ref{4.19}) and $\alpha(r)$ with its outer edge value given by formula (\ref{outer}) to obtain 
\be\label{4.25}
\tau\approx\left(\frac{\Gamma+1}{2}\right)
\gamma_\infty^2 r_o^{2\nu}(r-r_+)\;, \quad
\tau_{o}\approx
\left(\frac{\Gamma+1}{3}\right)\gamma_\infty^2 r_o^{1+2\nu}\epsilon
\; ; 
\ee
(to lowest order in $\epsilon$).

We now need to solve the radial equation (\ref{3.6}) in the capture zone $0<\tau<\tau_{o}$ with the boundary condition (\ref{3.12}) at $\tau=\tau_{o}$, and the requirement that the solution decays inwards through the disk for $\tau<0$. By its definition (\ref{3.6}) and formula (\ref{4.25}), the coefficient in the radial equation simplifies  to 
\be\label{4.26}
S(\tau)\approx s_+\tau \;,\qquad  s_+=\frac{dS}{d\tau}\Biggl|_{\tau=0}\approx
\frac{12}{\left(\Gamma+1\right)^2\,\gamma_\infty^4\, r_o^{4(1+\nu)}}\; . 
\ee\label{4.27}
Hence the radial equation (\ref{3.6}) becomes the Airy equation
\be\label{4.28}
\frac{d^2W}{d\tau^2}+s_+\tau\, W=0\; ,\qquad 0<\tau<\tau_{o}\; ,
\ee
whose solution satisfying the decay condition is the Airy function of the first kind,
\be
W \propto Ai(-s_+^{1/3} \tau)\; .
\ee
Substitution of this eigenfunction into the boundary condition (\ref{3.12}), using equations (\ref{4.23}) and (\ref{4.26}), provides the eigenfrequency equation in the form 
\be
\frac{2\epsilon}{r_o^{3}}\cos\theta_o Ai(-z)-s_+^{1/3}\sin\theta_o Ai^{'}(-z)=0\; ,\quad
z\equiv s_+^{1/3}\tau_{o}\approx\left[\frac{4\left(\Gamma+1\right)}{9}\right]^{1/3}\,
\left[\frac{\gamma_\infty}{r_o^{1/2-\nu}}\right]^{2/3}{\epsilon}\; .
\label{4.29}
\ee
(A prime denotes the derivative in the whole argument.) Unless $\theta_o$ is very close to zero, the first term here can be dropped, 
giving
\[
Ai^{'}(-z)=0\; ,\qquad \theta_o\not=0\; ;
\]
while in the opposite case
\[
Ai(-z)=0\; ,\qquad \theta_o=0\; .
\]
Both equations allow for an infinite sequence of positive roots 
$0<z_0<z_1<\dots<z_n<\dots$ which provide the eigenfrequencies according to equations (\ref{4.23}) and (\ref{4.29}), namely:
\[
\sigma_n=r_o^{-3/2}(1+\epsilon_n)\; , 
\]
\be
\epsilon_n=z_n \left[\frac{4\left(\Gamma+1\right)}{9}\right]^{-1/3}
\left[\frac{\gamma_\infty}{r_o^{1/2-\nu}}\right]^{-2/3}=
z_n \left[\frac{9}{4\left(\Gamma+1\right)}\right]^{1/3}
c_s^{2/3}(r_o,0)\,r_o^{1/3}\; .
\label{eigf}
\ee
Our requirement that $\epsilon_n$ be small enough (at least $\epsilon_n\ll 4$) can be easily met by its expression (\ref{eigf}), since the last factor there is small due to inequality (\ref{3.10}). Since for large $n$ the roots $z_n$ grow as 
\be
z_n\approx \left(3n\pi/2\right)^{2/3}, \qquad n\gg1\; ,
\label{4.39}
\ee
the largest possible radial mode number $N$ is found from
\be
N\ll (2/3\pi)\,\left[\frac{4\left(\Gamma+1\right)}{9}\right]^{1/2}
\left[\frac{\gamma_\infty}{r_o^{1/2-\nu}}\right]
\sim\frac{1}{c_s(r_o,0)r_o^{1/2}}\sim\frac{r_o}{h(r_o)}\; ,
\label{Nout}
\ee
and can thus be rather large.
 
Let us briefly summarize the physical properties of these outer LF fundamental p--modes. More quantitative information about them is given in section 8.

1. The spectrum of the outer LF fundamental p--modes is finite, although the total number of them can be large. 

2. All the eigenfrequencies are only slightly greater than
$r_o^{-3/2}$, so that they are practically independent of all other parameters, including the angular momentum parameter $a$. They have a rather weak dependence on the midplane speed of sound at the outer edge of the disk.

3. The radial extent of these modes, $r_o-(r_+)_n=(2/3)\epsilon_n r_o$, could be large compared to $r_i$.

\section{Inner Low-Frequency Fundamental P--Modes}

\subsection{Radial eigenvalue problem}

For $m=0$ and low frequency $\sigma$ the outer boundary ($r=r_-$) of the inner p--mode capture zone is close to the inner boundary (the inner edge $r=r_i$ of the disk). Thus it is natural to consider the eigenfrequency low if the linear approximation for the radial epicyclic frequency,
\be\label{4.30}
\kappa^2(r)\approx q^2(r-r_i) \; , \qquad 
q^2\equiv\frac{d\kappa^2}{dr}(r_i)=
6\left[\frac{r_i^{1/2}-a}{r_i^{3/2}\left(r_i^{3/2}+a\right)}\right]^2<1\; ,
\ee
is valid throughout the whole trapping zone. In particular, at its outer boundary
\be\label{4.31}
\sigma^2=\kappa^2(r_-)\approx q^2(r_--r_i) \; , \qquad r_-\approx r_i+\sigma^2/q^2\; .
\ee
Since the second derivative of $\kappa^2(r)$ is of the order of the first ($=q^2$), the condition for the linear approximation (\ref{4.30}) to hold is 
\be\label{4.32}
\sigma^2/q^2 \lesssim 1\; .
\ee
That is our definition of the low frequency range for this case. In contrast with the inequality (\ref{4.32}), the condition (\ref{4.18}) needed for the validity of the approximation (\ref{4.19}) for the vertical eigenvalue is easily met. It does not impose any serious constraints on the eigenfrequency, such as the lower limit (\ref{4.22}) in the case of the outer modes, at least when $-1\leq a\leq a_\perp\approx 0.953$. The function $q(a)$ is plotted in Fig. 1. It increases monotonically within this whole range of the angular momentum parameter, from the minimum value $q(-1)\approx 0.014$ up to the maximum $q(a_\perp)\approx 0.111$.

We now fix the variable $\tau$ from equation (\ref{3.5}) as
\be
\tau=\int_{r_i}^r\alpha^2(r^{'})
\left[1 - \frac{\Psi(r^{'})}{\omega_*^2(r^{'})}\right]
\,dr^{'}\;; \quad
\tau_{i}=0\;,\quad
\tau_-=\int_{r_i}^{r_-}\alpha^2(r^{'})
\left[1 - \frac{\Psi(r^{'})}{\omega_*^2(r^{'})}\right]
\,dr^{'} \; . \label{4.33}
\ee
Then from equations (\ref{4.19}), (\ref{inner}), and (\ref{4.31}) we obtain
\be
\tau\approx\frac{\left(\Gamma+1\right)\alpha_i^2}{2(1-2\mu)}(r-r_i)^{1-2\mu}\;,\quad
\tau_-\approx\frac{\left(\Gamma+1\right)\alpha_i^2}{2(1-2\mu)}
\left(\frac{\sigma}{q}\right)^{2(1-2\mu)}\; . \label{4.34}
\ee
Calculating the coefficient $S(\tau)$ in the radial equation (\ref{3.6}) using all the above approximations as well as the representation (\ref{inner}) for $\alpha(r)$ near the inner edge, we obtain
\be
\frac{d^2W}{d\tau^2}+b^2\tau^{2 \mu / (1 - 2\mu)}
\left[\tau_-^{1/(1-2 \mu)} - \tau^{1/(1-2 \mu)}\right]W=0,\quad
0<\tau<\tau_- \; , \label{4.35}
\ee
where
\be
b^2=b^2(\mu) \equiv (1 - 2 \mu)^{(1 + 2 \mu) / (1 - 2\mu)}q^2
\left[2/(\Gamma + 1) \alpha_i^2 \right]^{2/(1-2\mu)} \; . \label{4.36}
\ee
Equation (\ref{4.35}) should be solved with the boundary condition (\ref{3.11}) and the condition of decay outside $\tau_-$. Unfortunately, it cannot be integrated in terms of the known special functions, so we treat it asymptotically with respect to the parameter $0\leq\mu<1/2$.

\subsection{Solution for a weak inner edge singularity ($\mu\to+0$)}

We observe that for disks with a non-singular or weakly singular inner edge, to lowest order in $\mu\ll 1/2$ equation (\ref{4.35}) becomes the Airy equation
\[
\frac{d^2W}{d\tau^2}+b^2_0\left(\tau_- - \tau\right)W=0,\quad
b^2_0\equiv b^2(0)= \left[2q/(\Gamma + 1) \alpha_i^2\right]^{2}
\;.
\]
Its solution satisfying the decay condition is $W \propto Ai[-b_0^{2/3}(\tau_- - \tau)]$. In complete similarity to the previous section, the boundary condition produces
\be
\sigma_n = z_n^{1/2}[2/(\Gamma + 1)]^{1/6}q^{2/3}\alpha_i^{-1/3},
\qquad n=0,1,2,\dots\; ,
\label{4.37}
\ee
where $z_n$ is the $(n+1)$-th positive root of either $Ai(-z)$ or $Ai^{'}(-z)$, depending on whether the function or its derivative has a larger coefficient in the boundary condition (\ref{3.11}) (i.~e., on how $\sigma^2$ compares to $b^{2/3}_0$). 
The LF inequality (\ref{4.32}) applied to the lowest mode from equation (\ref{4.37}) requires that
\be
(q \alpha_i)^{2/3} \gg 1\; . \label{4.38}
\ee
This is true for typical values of the parameters involved, so generically LF fundamental inner modes in the disks with a weak inner edge singularity do exist. In addition, the condition (\ref{4.38}) implies that $\sigma_n^2\gg b^{2/3}_0$; therefore $z_n$ is the root of the Airy function $Ai(-z)$ 
(unless $\theta_i = \pi/2$).

For large enough values of $n$ the asymptotic behavior (\ref{4.39}) for $z_n$ holds, 
while on the other hand the LF condition (\ref{4.32}) must still be valid. Hence the largest possible radial mode number $N$ is determined by
\be
 N \ll (2q \alpha_i/3\pi)[(\Gamma + 1)/2]^{1/2}\; ,
\label{4.40}
\ee
for $N \gtrsim 5$. 

A physically meaningful way to represent the eigenfrequency  is
\be
\sigma_n=
z_n^{3/2}\left[2/(\Gamma + 1)\right]^{1/2}c_s(r_-,0)/l_n \; , \label{sphys1}
\ee
where $r_-=(r_-)_n$, and $l_n=(r_-)_n-r_i$ is the length of the mode capture zone. One derives this from equation (\ref{4.37}) by removing $q$ with the help of equation (\ref{4.31}), $q^2=\sigma_n^2/l_n$, and also noticing that for $\mu\approx0$ and low frequency, the speed of sound $c_s(r,0)\approx\alpha_i^{-1}$ throughout most of the capture zone.

\subsection{Solution for a strong inner edge singularity }

The limit case of a disk with a strong inner edge singularity, $\mu\approx 1/2-0$, is apparently a singular one for equation (\ref{4.35}). To investigate it, we once again change the independent variable to
\be
x\equiv(\tau/\tau_-)^{1/(1-2\mu)}\;
\label{4.41}
\ee
to obtain
\be
\frac{d}{dx}\left(x^{2\mu}\frac{dW}{dx}\right)+\zeta^2(1-x)W=0 \; , \qquad
0<x<1 \; , \label{4.42}
\ee
with
\be
\zeta^2=\zeta^2(\mu,\sigma) \equiv 
0.5\left(\Gamma + 1\right)\alpha_i^2 q^{ -4(1-\mu)} \sigma^{ 2(3 - 2\mu)}\; . \label{4.43}
\ee
To lowest order in $1-2\mu\to+0$, we can replace $x^{2\mu}$ with $x$ in the equation (\ref{4.42}). Then we also make a substitution
\[
W(x)=e^{-\zeta x}w(x) \; ,
\]
reducing it to
\[
x\,\frac{d^2w}{dx^2}+(1-2\zeta x)\,\frac{dw}{dx}-\zeta(1-\zeta)\,w=0 \; , \qquad 0<x<1\; .
\]
Up to a rescaling of the independent variable, this is the confluent hypergeometric equation, whose only solution non-singular at $x=0$ is the Kummer function (confluent hypergeometric function of the first kind).

Hence
\[
W(x)= e^{-\zeta x}\,\Phi\left(\frac{1-\zeta}{2},\,1,\,2\zeta x\right)\sim
\frac{(2\zeta x)^{-(1+\zeta)/2}}{\Gamma_E[(1-\zeta)/2]}\,e^{\zeta x}\,
\to\,\infty \; , \quad x\to\infty\; ,
\]
unless $(1-\zeta)/2=-n,\quad n=0,1,2,...$ [$\Gamma_E(z)$ is the Euler gamma function]. The condition for the mode to decay on the right of $x=1$ can be thus fulfilled in the latter case only. The Kummer function then becomes the Laguerre polynomial $L_n$, and the eigenvalue and its eigenfunction become
\be
\zeta_n=2n+1\; ,\qquad W_n(x)= e^{-(2n+1)x}\,L_n(2(2n+1)x)\; ,\qquad n=0,1,2,...\; . 
\label{4.44}
\ee
The eigenfrequency is found from equations (\ref{4.44}) and (\ref{4.43}) to be
\be
\sigma_n=
\left\{
(2n+1)\left[2/(\Gamma + 1)\right]^{1/2}q^{2(1-\mu)}\alpha_i^{-1}
\right\}^{1/(3-2\mu)}\; .
\label{4.45}
\ee
As in the previous case, for the inner LF fundamental p--modes in the disk with a strong singularity to exist at all, condition (\ref{4.32}) should hold at least with $\sigma_0$ from equation (\ref{4.45}). This implies that
\be
(q \alpha_i)^{2/(3-2\mu)} \gg 1\; ,
\label{ex2}
\ee 
which is just slightly different from the corresponding inequality (\ref{4.38}), and typically is also valid. The maximum admissible mode number $N$ is determined by
\be
 N \ll \,0.5\left[(\Gamma + 1)/2\right]^{1/2}q\alpha_i\; ,
\label{4.47}
\ee
and its value can be large enough, comparable to the number of inner modes found for a weakly singular disk given by equation (\ref{4.40}), which is about two times smaller.

Note that the expression (\ref{4.45}) for the eigenfrequency to lowest order in $(0.5-\mu)\approx0$ simplifies to
\be
\sigma_n=
\left\{
(2n+1)\left[2/(\Gamma + 1)\right]^{1/2}(q/\alpha_i)
\right\}^{1/2}\; .
\label{mu1/2}
\ee
In terms of physical parameters, that is the speed of sound and the trapping zone length, the eigenfrequency can also be written as 
\be
\sigma_n=
(2n+1)\left[2/(\Gamma + 1)\right]^{1/2}c_s(r_-,0)/l_n \; , \label{sphys2}
\ee
where, as before, $r_-=(r_-)_n$ and $l_n=(r_-)_n-r_i$. 
In this form it is very similar to the eigenfrequency (\ref{sphys1}) for the previous case, the only difference being the numerical factor $(2n+1)$ instead of $z_n^{3/2}$.

In addition, if we formally set $\mu=0$ in the expression (\ref{4.45}) for the eigenfrequency [which, in fact, is valid in the opposite limit case ($\mu\to 0.5-0$)], we obtain
\[
\sigma_n = (2n+1)^{1/3}[2/(\Gamma + 1)]^{1/6}q^{2/3}\alpha_i^{-1/3}\; , \quad n=0,1,2...\; .
\]
Quite remarkably, this expression essentially coincides with the `small $\mu$ formula' (\ref{4.37}), with the exception of the first factor giving the dependence on the mode number [$(2n+1)^{1/3}$ in place of $z_n^{1/2}$]. For the lowest mode ($n=0$), even these factors differ by only about 5\%.

Finally, in both limit cases of the weak and strong edge singularity, it is not difficult (although hardly needed) to calculate higher order corrections using an expansion of the type
\[
u^{\epsilon}= \exp(\epsilon\ln u)=1+\epsilon\ln u+0.5\,(\epsilon\ln u)^2+... \; ,\quad 
\epsilon\to0\; ,
\]
with $u=\tau,\,\, \epsilon= 2\mu/(1-2\mu)$ for $\mu\approx0$; and $u=x,\,\, \epsilon= 1-2\mu$ for $\mu\approx1/2-0$. The weak logarithmic singularity at $u=0$ does not prevent the use of standard perturbation techniques, as carried out in section 3.

\section{LF Fundamental P--modes vs. Fundamental C--modes}

It is both natural and useful to discuss the features of the LF fundamental p--modes by comparing them to the LF fundamental c--modes found in \citet{swo}. For brevity, in this section we speak about just p-- or c--modes, always meaning only low frequency fundamental modes.

\subsection{Eigenfrequency spectrum}

{\it a) Angular and vertical mode numbers.} They are $m=j=0$ for p--modes and $m^2=j=1$ for c--modes.

{\it b) Radial mode number.} In both cases, the number of the radial modes ($N$) is finite. Therefore, the whole eigenfrequency spectrum is finite.

\subsection{Spectrum dependence on parameters}

{\it a) Dependence on the angular momentum.} The  p--modes exist within the range $-1\leq a\leq a_\perp\approx 0.953$. The c--modes exist only for corotating disks, in the range $0<a_0\leq a\leq a_\perp$. The cutoff value $a_0\sim10^{-5}-10^{-3} $ depends on the adiabatic index $\Gamma$, the behavior of the speed of sound through the whole capture zone (in particular, on both $\mu$ and $\nu$), and slightly on the boundary condition parameter $\theta_i$.

For the outer p--modes the dependence of eigenfrequencies on $a$ is very weak (through the value of the speed of sound at the outer edge only). For the inner p--modes it is essential, but for c--modes it is very sharp, especially near the cutoff value $a_n$ for the mode number $n$ (with $a_n>a_{n-1}>\dots>a_1>a_0$). The total number of the modes, $N$, depends on all the parameters, but changes most strongly with $a$ in the case of c--modes.

{\it b) Dependence on the boundary condition parameter $\theta_i$.} The dependence of p--modes on $\theta_i$ is very weak. The dependence of c--modes on the boundary condition parameter is also not strong.

{\it c) Dependence on the speed of sound }(i.~e., on $\mu$, $\nu$, $\alpha_i$, $\gamma_\infty$). The p--modes depend essentially on the behavior of the speed of sound near either the inner or outer edge of the disk; that is, on $\mu$ and $\alpha_i$ (inner modes) or $\nu$ and $\gamma_\infty$ (outer modes). The c--modes depend on all the parameters, including a considerable dependence on $\nu$. 

{\it d) Lower bound for the eigenfrequencies}. The eigenfrequencies are bounded from below in all cases: $|\sigma_n| > [2/(\Gamma + 1)]^{1/6}q^{2/3}\alpha_i^{-1/3}$ for inner p--modes, $|\sigma_n| > r_o^{-3/2}$ for outer p--modes, and $|\sigma_n| \geq 2a_0 r_o^{-3}$ for c--modes. Note that the lower bound for c--modes depends on the angular momentum parameter and the outer radius of the disk (in fact, the bound is the Lense--Thirring frequency at the outer edge). For inner p--modes, the bound depends on the angular momentum parameter, the behavior of the speed of sound at the inner edge, and weakly on the adiabatic index $\Gamma$. For outer p--modes the only dependence is on the outer radius.

\subsection{Mode capture zones}

The capture zone of both inner and outer p--modes shrinks as the eigenfrequency approaches its lower limit. The radial extent of the capture zone of the c--modes increases as their eigenfrequencies decrease, and can cover the whole disk. The difference in the behavior is mainly due to the difference in the angular modes numbers, $m=0$ and $m=\pm1$, respectively.
[See Figure 1 in \cite{swo}.]

\subsection{Oscillation forms}

To lowest order, for p--modes
\[
\delta p^{(p)}/\rho\,\propto\,\delta V^{(p)}\,\propto\, 
f_n^{(p)}(r)\exp[i\sigma_n^{(p)} t] \; ,
\]
while for c--modes
\[
\delta p^{(c)}/\rho\,\propto\,\delta V^{(c)}\,\propto\,
 zf_n^{(c)}(r)\exp[i(\pm\phi+\sigma_n^{(c)} t)]\; .
\]
Therefore p--modes are almost purely radial, since the oscillations of particles in the vertical direction are suppressed,
\[
\xi^z\,\propto\,[\partial(\delta V^{(p)})/\partial z]\,\approx\,0 \; .
\]
[$\xi^z$ is the vertical component of the Lagrangian displacement from equilibrium \citep{per}.] The vertical oscillations of the particles in a c--mode are independent of $z$. Particles on the midplane do not oscillate in the radial direction, 
\[
\xi^r\,\propto\,[\partial(\delta V^{(c)})/\partial r]\,\approx\,0\quad{\rm at}\quad z=0\; .
\]

\section{Numerical Results and Discussion}

\subsection{The unperturbed model}
  
We now apply the above analysis to models of black hole accretion disks.
We start by stating our assumptions about the unperturbed model, in addition to what has already been said in section 2. The results discussed below were obtained by assuming that gas pressure dominates in the disk, so that $\Gamma = 5/3$, and that the opacity is due to (optically thick) electron scattering. These conditions hold, regardless of the luminosity, at least near the inner edge $r_i$ and for $r \gg r_i$. [For example, for $L/L_{Edd} = 0.1$, $M/M_\sun = 10$, and (the usual viscosity parameter) $\alphav = 0.1$, 
the inner region where gas pressure dominates extends from $r_i$ to $1.15 r_i$ for 
$a = 0$, and from $r_i$ to $1.079 r_i$ for $a = 0.953$.]

It is notable that the interior structure of the (zero buoyancy) accretion disk
enters our formulation only through the constant $\Gamma$ and the function
$\alpha(r)$, defined by equation (\ref{defalpha}). 
We modeled $\alpha(r)$ in the way described by equation (\ref{3.7}).
The relativistic factor 
\be
\beta_{rel} \sqrt{g_{rr}} =  (1+a r^{-3/2}) [(1-3r^{-1}+2a r^{-3/2})
  (1 - 2 r^{-1} + a^2 r^{-2})]^{-1/2}
\ee
that appears in the definition of $\alpha(r)$ is of order unity except for values of $a$ close to 1 and simultaneously $r$ close to $r_i \approx 1$. It is shown in Figure 2.

The values for the speed of sound are derived from the fully relativistic results of \citet{nt}:
\begin{eqnarray} 
c_s/c & = & 8.61 \times 10^{-3} (L/L_{Edd})^{1/5} (\alphav M/M_\sun)^{-1/10}
\eta^{-1/5}(a)r^{-9/20} \; , \quad r \gg r_i \; , \label{spe1} \\ 
c_s/c & = & 8.61 \times 10^{-3} (L/L_{Edd})^{1/5}(\alphav M/M_\sun)^{-1/10} 
r_i^{-9/20}\eta^{-1/5}(a) \nonumber \\
& &  \times {\cal B}^{-1/5}(a,r_i) {\cal D}^{-1/10}(a,r_i)
f^{1/5}(a) (r-r_i)^{2/5} \; , \quad r \cong r_i, \mu > 0 \; , \label{spe2} \\
c_s/c & = & 8.61 \times 10^{-3} (L/L_{Edd})^{1/5} 
(\alphav M/M_\sun)^{-1/10}r_i^{-9/20}\eta^{-1/5}(a) \nonumber \\ 
& & \times {\cal B}^{-1/5}(a,r_i) {\cal D}^{-1/10}(a,r_i)  
F^{1/5}(a) (1 - \beta)^{1/5} \; , \quad r \cong r_i, \mu = 0 \; . \label{spe3}
\end{eqnarray}
(As we mentioned above, the cases $\mu > 0$ and $\mu = 0$ correspond
to the absence and presence, respectively,
of torque at the inner edge of the disk.)
In these expressions, $\eta$ (shown in Figure 3) is the `efficiency factor' which relates the mass accretion rate to the luminosity: $L = \eta \dot{M}c^2$. (It is the energy lost
by a unit mass as it accretes from far away to the inner edge.)
The constant $\beta$ is the fraction of the inflowing specific angular momentum at $r_i$
that is absorbed by the black hole (so there is a torque at $r_i$ unless $\beta = 1$).
The functions ${\cal B}$ and ${\cal D}$ are relativistic correction factors defined in \citet{nt}, which approach unity at large $r$. The function $f$ (shown in Figure 4) is defined such that ${\cal Q} = f(a)(r - r_i)^2$ to lowest order in $r-r_i$, where ${\cal Q}$ [defined in \citet{nt,pt}] is the factor responsible for making $\alpha(r)$ singular at $r_i$.
In the presence of a torque at $r_i$, ${\cal Q}(r_i) = F(a)(1-\beta)$, where 
\be
F(a) \equiv (1 - 2 r_i^{-1} + a^2 r_i^{-2})^{1/2}
(1 - 2 a r_i^{-3/2} + a^2 r_i^{-2}) (1 - 3 r_i^{-1} + 2 a r_i^{-3/2})^{-1/2} \; .
\ee
This function is shown in Figure 5.

By comparing expressions (\ref{3.7}) and (\ref{spe1}), one obtains $\nu = 9/20$ (which we employ for all cases). Equation (\ref{spe2}) yields $\mu = 2/5$ for the case
of no torque at the inner edge, while equation (\ref{spe3}) is consistent with $\mu = 0$ for the case of a nonvanishing torque at the inner edge.

Moreover, we are now ready to determine the actual parameters involved in the eigenfrequencies, that is, $\alpha_i$ and $\gamma_\infty$. According to the definitions (\ref{alpha}) and (\ref{inner}),
\[
\alpha_i=\lim_{r\to r_i+0}(r_i-r)^{\mu}\beta_{rel}\sqrt{g_{rr}}/c_s(r,0) \; .
\]
By employing equations (\ref{spe2}) and (\ref{spe3}), we then obtain
\begin{eqnarray}
\alpha_i(a,M,L/L_{Edd},\alpha_*) &=& 
116 (L/L_{Edd})^{-1/5}(\alphav M/M_\sun)^{1/10} 
(\beta_{rel} \sqrt{g_{rr}})_{r_i}
r_i^{9/20} 
\eta^{1/5}(a) \nonumber \\ & & \times
{\cal B}^{1/5}(a,r_i) 
{\cal D}^{1/10}(a,r_i)
f^{-1/5}(a),
\qquad \mu =2/5 \; , \label{alphamu2}
\end{eqnarray}
and
\begin{eqnarray}
\alpha_i(a,M,L/L_{Edd},\alpha_*) &=&
116 (L/L_{Edd})^{-1/5}(\alphav M/M_\sun)^{1/10} 
(\beta_{rel} \sqrt{g_{rr}})_{r_i}
r_i^{9/20} 
\eta^{1/5}(a) \nonumber \\
& & \times
{\cal B}^{1/5}(a,r_i) 
{\cal D}^{1/10}(a,r_i)
F^{-1/5}(a) (1-\beta)^{-1/5},
\qquad \mu =0 \; . \label{alphamu0}
\end{eqnarray} 
These two functions are shown in Figures 6 and 7. 
We see that in both cases $\alpha_i$ has a value which is 
approximately constant except 
for $a \gtrsim 0.7.$
These properties are reflected in the properties of inner fundamental LF p--modes described below.

Similarly, from equations (\ref{alpha}) and (\ref{outer}) we obtain
\[
\gamma_\infty=\lim_{r\to \infty}r^{-\nu}\beta_{rel}\sqrt{g_{rr}}/c_s(r,0) \; ,
\]
which, in view of equation (\ref{spe1}), gives $(\nu=9/20)$:
\be
\gamma_\infty(a,M,L/L_{Edd},\alpha_*) = 
116 (L/L_{Edd})^{-1/5}(\alphav M/M_\sun)^{1/10} 
(\beta_{rel} \sqrt{g_{rr}})_{r_i}
\eta^{1/5}(a)\; . \label{gamma}
\ee
This function (which determines the properties of outer fundamental LF p--modes) is shown in Figure 8.

Typical values for disk density and central object mass make the neglect of self-gravity a common assumption. An order-of-magnitude analysis shows that self-gravitational 
effects can be ignored, either at the unperturbed or the perturbed level
(Cowling approximation), whenever
\begin{equation}\label{sg-per}
 \frac{r}{GM/c^2} \ll \frac{10^{11}}{(M/M_{\odot})^{0.96}} \; ,
\end{equation}
in normal units. These estimates assume $\alphav = 0.1$ and $L = L_{Edd}$, with this condition (\ref{sg-per}) being only slightly less restrictive for lower values of the luminosity.

Thus, for stellar mass black holes the condition is satisfied by many orders of magnitude, while one needs to be more careful with supermassive black holes when studying the outer modes. A black hole mass of $10^8 M_\sun$, for example, implies the constraint
$r \ll 2\times 10^3 GM/c^2$.

\subsection{Numerical results}

We present our major observationally relevant results in Figures 9--15 and Table 1,
obtained from the formulas derived in sections 5 and 6.
For the outer p--modes, we take $r_o = 10^4(GM/c^2)$ for the stellar case and $r_o = 10^3(GM/c^2)$ for the supermassive case (unless otherwise indicated), and $\theta_o\not=0$ in the boundary condition. For the figures, we have chosen the radial mode number $n = 0$. As with the g-- and c--modes, one might expect that the lowest radial mode would be the one most easily excited and with the largest net modulation. The viscous parameter is taken to be $\alphav = 0.1$ throughout.  

In Figure 9 we plot the fractional radial extent of the fundamental outer p--mode as a function of the angular momentum of the black hole. This quantity is $2\epsilon_0/3$ [as defined in equation (\ref{eigf})], so we see that the model is self-consistent in that $\epsilon_0 \ll 4$. The frequency is then essentially just $r_o^{-3/2}$.
Note that given the dependence of the speed of sound on $L$ and $M$ [expressions (\ref{spe1}), (\ref{spe2}) and (\ref{spe3})], all the p--mode plots will show the same `universal' characteristic: higher values of the frequency and fractional radial extent for higher values of $L$ and lower values of $M$.

Figure 10 shows the dependence of the eigenfrequency of the fundamental $\mu = 2/5$ (no torque) inner p--mode on $a$.
The fractional radial extent of this type of mode is shown in Figure 11. All of the modes satisfy condition (\ref{4.32}).

In Figure 12 we plot the eigenfrequency of the fundamental $\mu = 0$ inner p--mode as a function of $a$. Employing the results of \citet{hk}, a value of $\beta = 0.95$ has been chosen. The fractional radial extent of this type of mode appears in Figure 13. Note that these ($\mu = 0$) modes for which $M = 10 M_\sun$ and $L = L_{Edd}$ approach the limit of validity given by equations (\ref{4.31}) and (\ref{4.32}), since $\sigma_0^2/q^2 = 2.1(q\alpha_i)^{-2/3} = r_- - r_i\rightarrow 0.66$ as $a\rightarrow -1$. 
Comparing Figures 10 and 12, we notice that the 
frequencies are less sensitive to the value of 
$\mu$ than is the radial extent.

Figure 14 shows the
total number of radial 
modes allowable by our approach [which assumes equation (\ref{4.32})].
 

Another observationally relevant relation is that between the frequency and the luminosity. For the case of the inner p--modes, such a relation can be obtained analytically:
\be
d \log \sigma/d \log L = 1/15 \; ,
\ee
valid for both the $\mu\rightarrow 0$ and the $\mu\rightarrow 1/2$ cases.

Table 1 shows the frequency of the first three excited radial modes relative to the fundamental one. 
For the inner modes, this ratio does not depend on $L$, $M$, $a$, or any other parameter except $\beta$, which we have taken to be 1.00 (corresponding to $\mu = 2/5$) and 0.95 (for $\mu = 0$). For the outer modes there is some dependence on the other parameters (the ranges of $L$ and $M$ are the same as in the figures).

\subsection{Discussion}

The frequencies of all the fundamental diskoseismic modes that have been studied are collected in Figure 15. They have been compared with observations of pairs of `stable' high frequency quasi-periodic oscillations in two low-mass X-ray binaries with black hole candidates by \citet{wso}. Identifying the oscillations with the g-- and c--modes produced possible values of the mass and angular momentum of the black hole.
The top curve in Figure 15 corresponds to the orbital frequency of a free `blob' at the inner edge of the accretion disk. 
Only the (more realistic) $\mu = 0$ inner p--modes have been included in the plot.
The outer p--mode curve represents an upper limit to the frequency of this type of mode. It corresponds to
$L = L_{Edd}, M = 10 M_\sun,$ and $r_o = 10^3 GM/c^2$.

One of the reasons that the (inner) p--mode was excluded as a source of the observed oscillations is its relatively narrow radial extent, as indicated in Figures 11 and 13. It lies very close to the inner edge of the disk, so has very little luminosity to modulate. In addition, the radial velocity within the equilibrium disk (which we have neglected) becomes significant near the inner edge. As indicated in section 1, we are now investigating higher frequency p--modes, which cover more of the disk. We hope to relate this work to the traveling p--waves that \citet{hmk} and \citet{mt} claim to be generated in their numerical simulations of viscous accretion disks.

The c--mode has a somewhat larger radial extent. The g--mode has the largest radial extent, and is the only high-frequency oscillation that does not extend to the inner edge of the disk. Given the uncertainties of the physical conditions there, it is the most robust mode (being trapped purely gravitationally). The viscous damping and growth of all three classes of modes has been investigated by \citet{ort}. However, attention has not been focused on the outer p--modes because of the uncertain physical conditions near the outer edge of the disk.

The inner p--mode could be an important source of modulation of the accretion flow. This could be most observable in the emission from the surface of (weakly magnetized) neutron stars in binary systems. 
In general, we might learn more about the uncertain conditions at the inner
edge of the disk via these modes.

In the future we hope to extend this perturbation analysis to more realistic models of accretion disks. It promises to be challenging, since recent numerical simulations [such as by \citet{hk}] indicate that a stationary equilibrium may not exist on the length scales of our modes. If so, one can still hope that its characteristic frequencies are less than those of our modes. We do plan to include the effects of radial velocity as well as greater disk thickness.

\acknowledgments

This work was supported by NASA grant NAS 8-39225 to Gravity Probe B.

\begin{deluxetable}{ccccccc} 
\tablecolumns{7} 
\tablewidth{0pc} 
\tablecaption{Value of $\sigma_n/\sigma_0$} 
\tablehead{ 
\colhead{} &
\colhead{} &
\colhead{outer} &
\colhead{} & 
\multicolumn{3}{c}{inner} \\ 
\cline{3-3}
\cline{5-7} \\ 
\colhead{$n$} &
\colhead{} & 
\colhead{} &
\colhead{} & 
\colhead{$\mu = 2/5$} &
\colhead{} & 
\colhead{$\mu = 0$}    }
\startdata 
0 & & 1.00 -- 1.00 & & 1.00 & & 1.00 \\
1 & & 1.02 -- 1.12 & & 1.65 & & 1.32 \\
2 & & 1.03 -- 1.20 & & 2.08 & & 1.53 \\
3 & & 1.04 -- 1.28 & & 2.42 & & 1.70 \\
\enddata 

\end{deluxetable}

\begin{figure}
\figurenum{1}
\epsscale{1}
\plotone{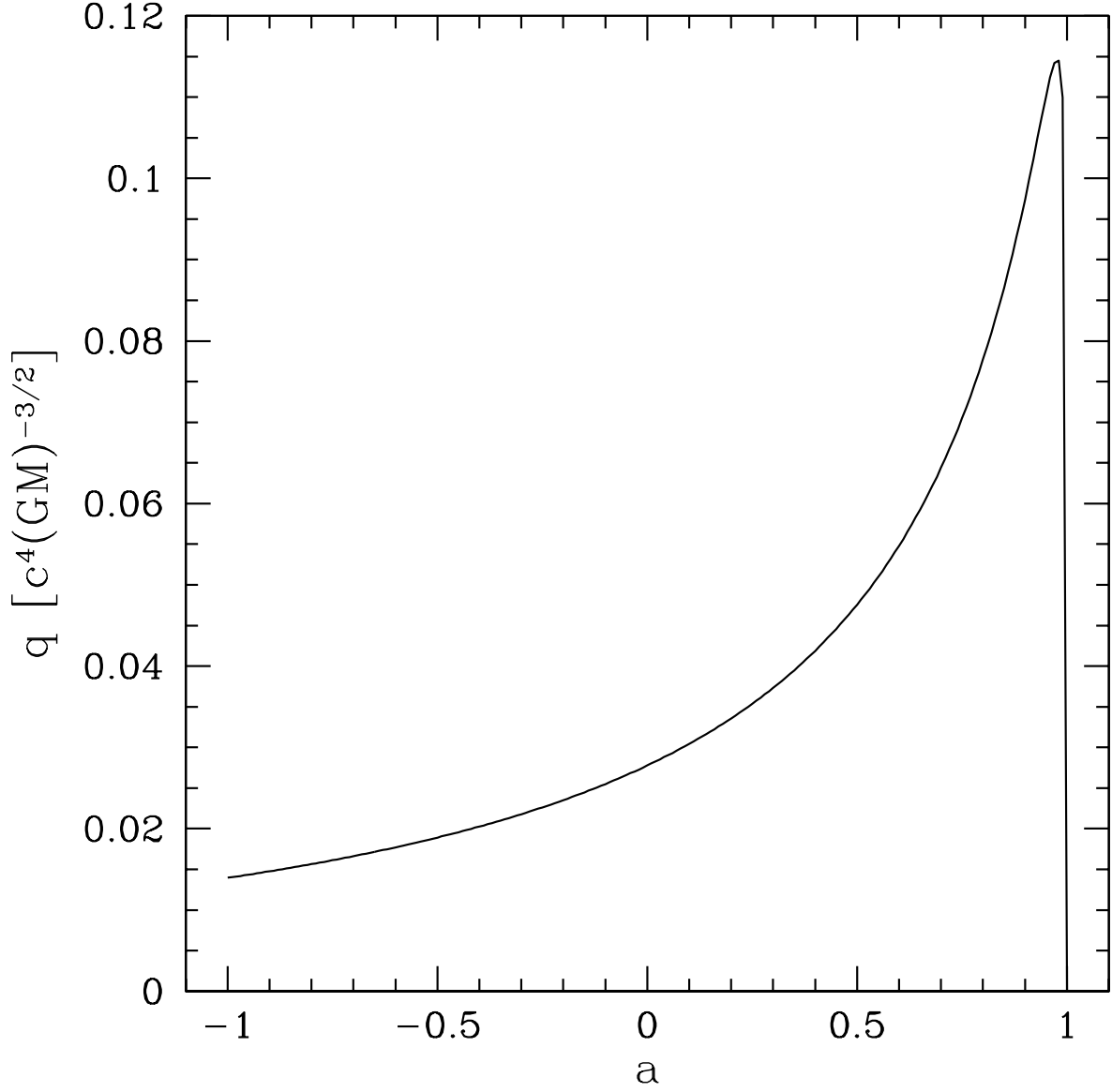}
\caption{Dependence of $q$,
defined by $q^2 \equiv (d\kappa^2/dr)_{r_i}$,
on the angular momentum of the black hole.}
\end{figure}

\begin{figure}
\figurenum{2}
\epsscale{1}
\plotone{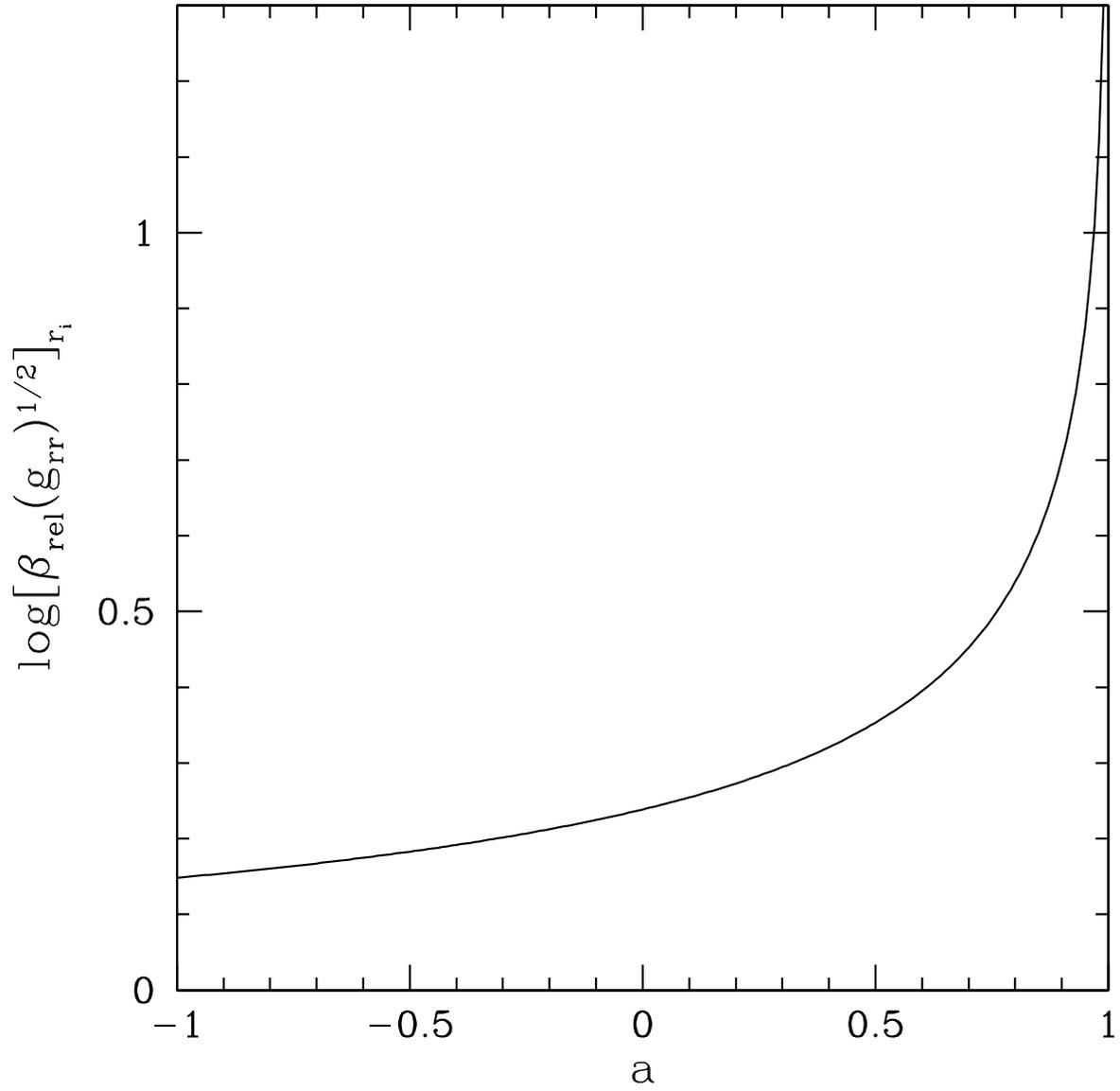}
\caption{Dependence of $(\beta_{rel} \sqrt{g_{rr}})_{r_i}$ on the angular momentum of the black hole.}
\end{figure}

\begin{figure}
\figurenum{3}
\epsscale{1}
\plotone{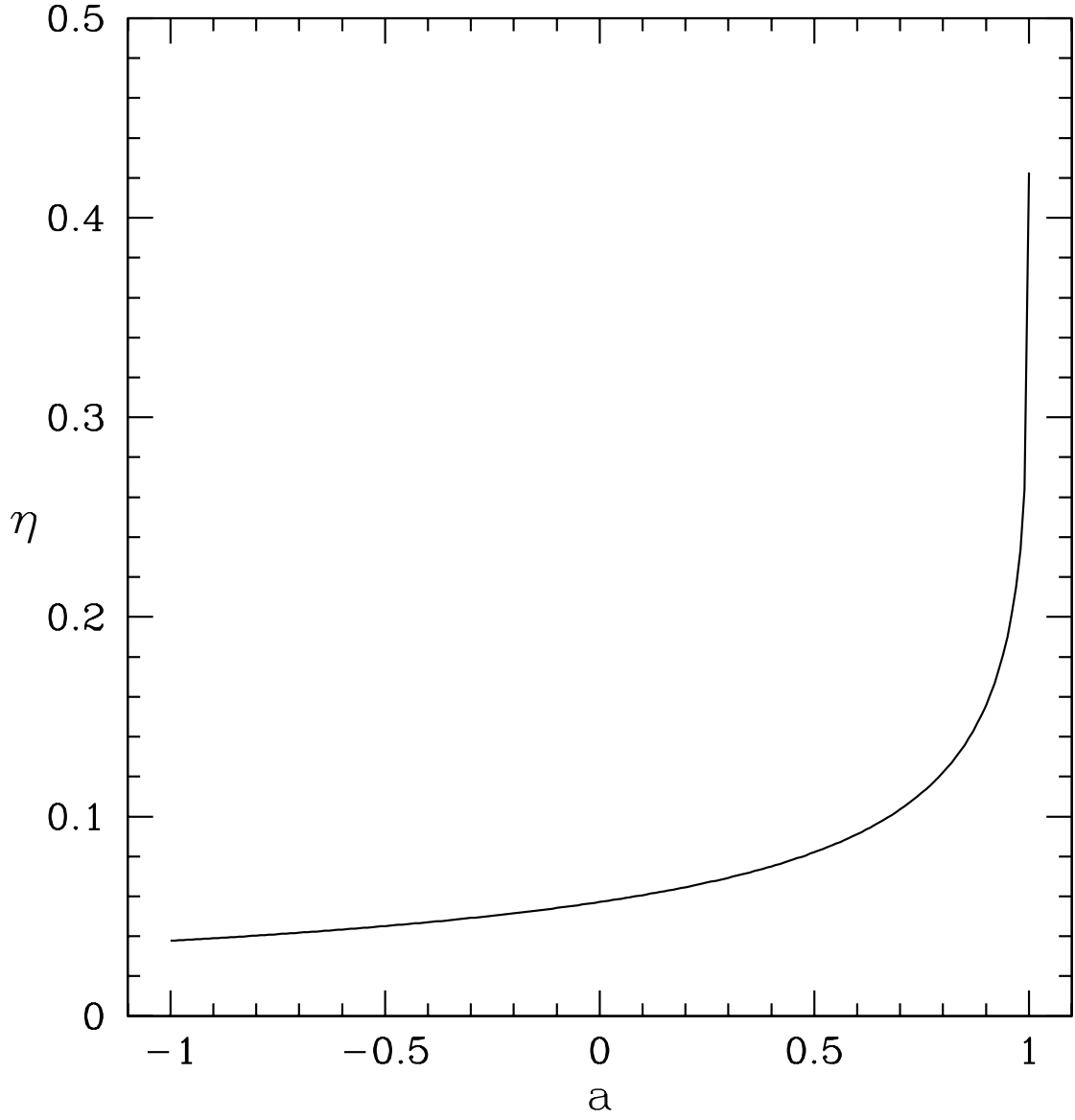}
\caption{Dependence of the efficiency factor $\eta$ on the angular momentum of the black hole.}
\end{figure}

\begin{figure}
\figurenum{4}
\epsscale{1}
\plotone{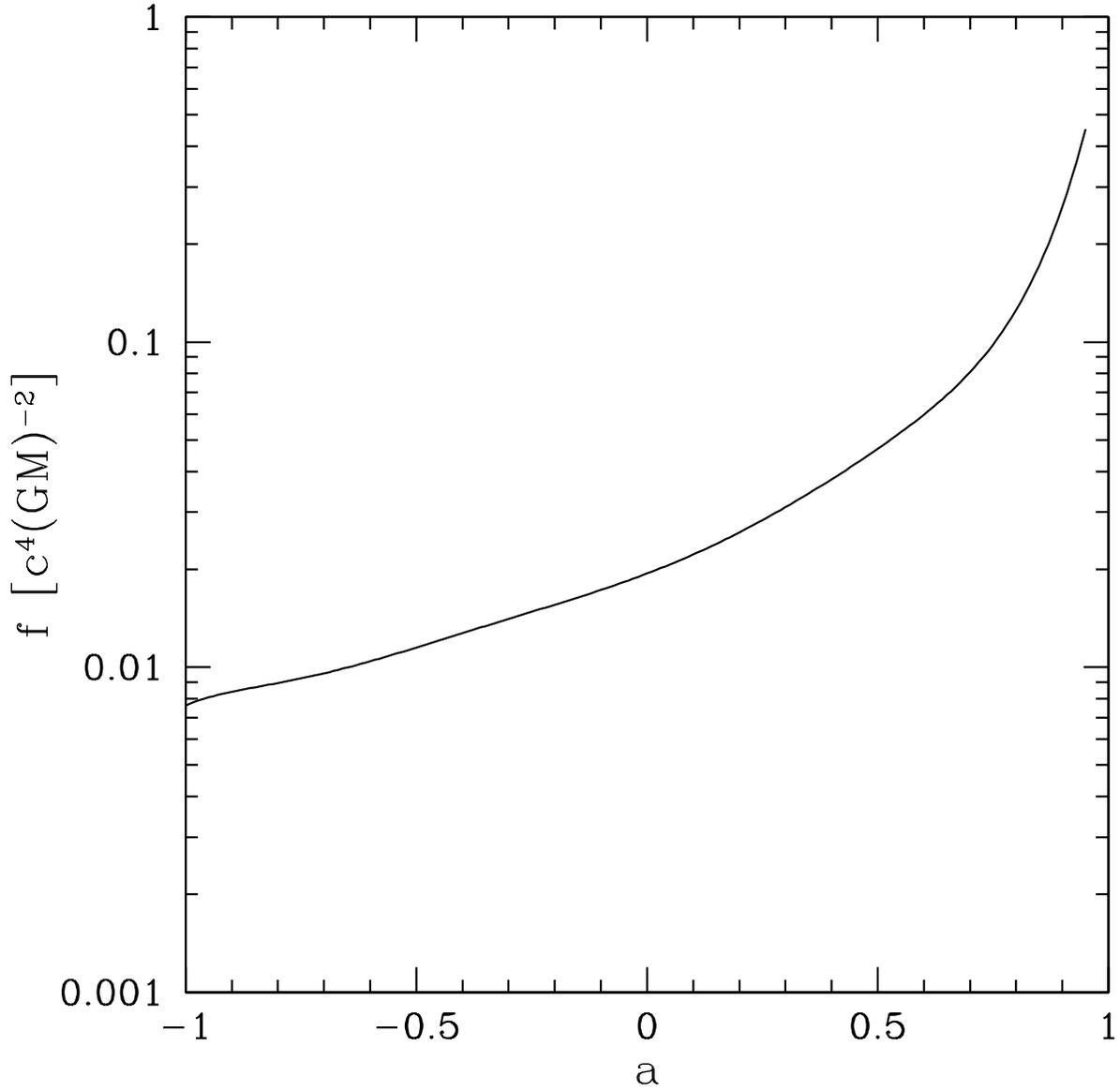}
\caption{Dependence of $f$ on the angular momentum of the black hole. It is defined in such a way that, for the case of no torque at $r_i$, ${\cal Q} = f(a)(r - r_i)^2$ to lowest order in $r-r_i$.}
\end{figure}

\begin{figure}
\figurenum{5}
\epsscale{1}
\plotone{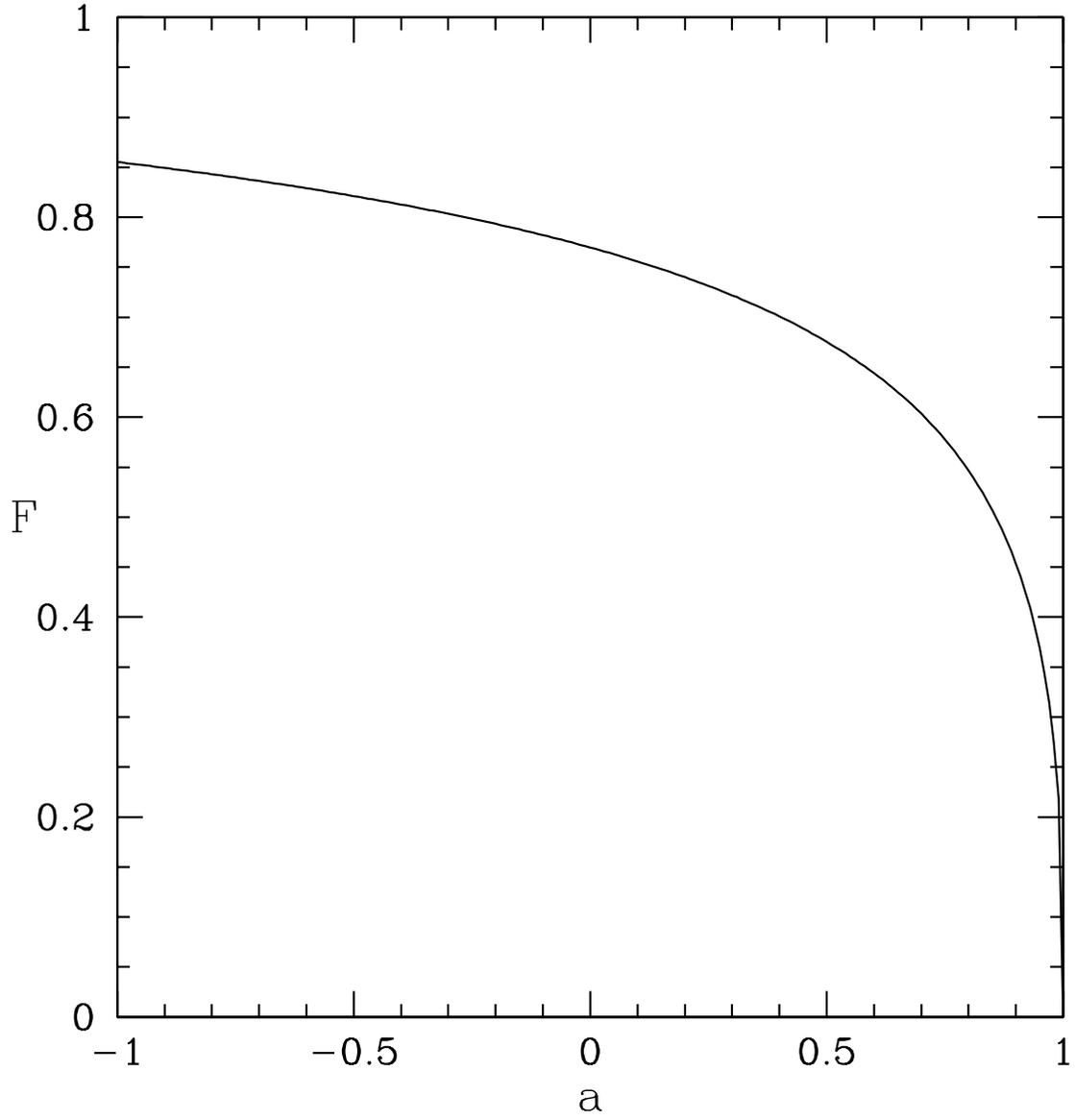}
\caption{Dependence of $F$ on the angular momentum of the black hole. 
It enters the relation ${\cal Q}(r_i) = F(a)(1-\beta)$.}
\end{figure}

\begin{figure}
\figurenum{6}
\epsscale{1}
\plotone{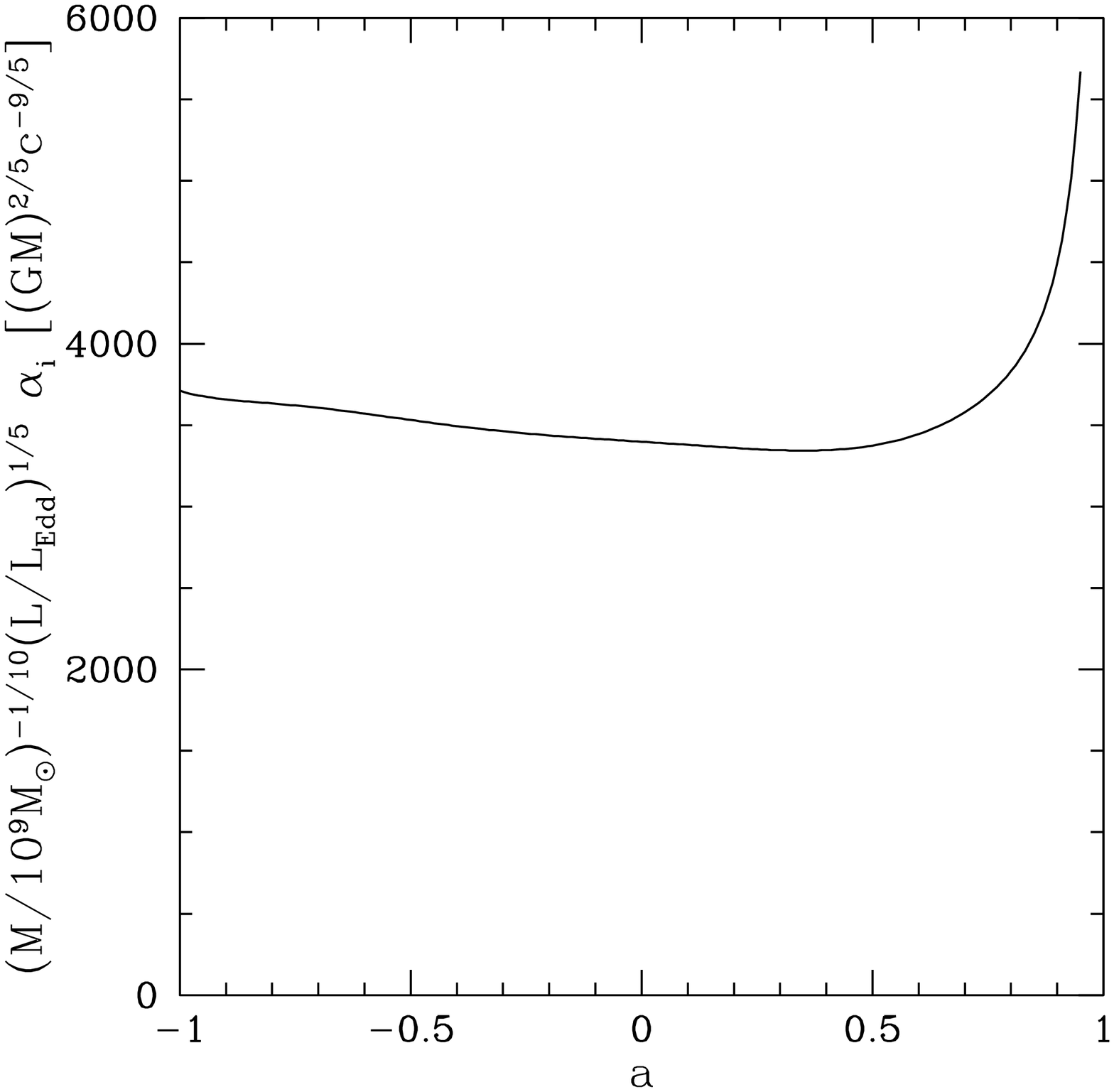}
\caption{Dependence of $\alpha_i$ on the angular momentum of the black hole, 
for the $\mu = 2/5$ case.}
\end{figure}

\begin{figure}
\figurenum{7}
\epsscale{1}
\plotone{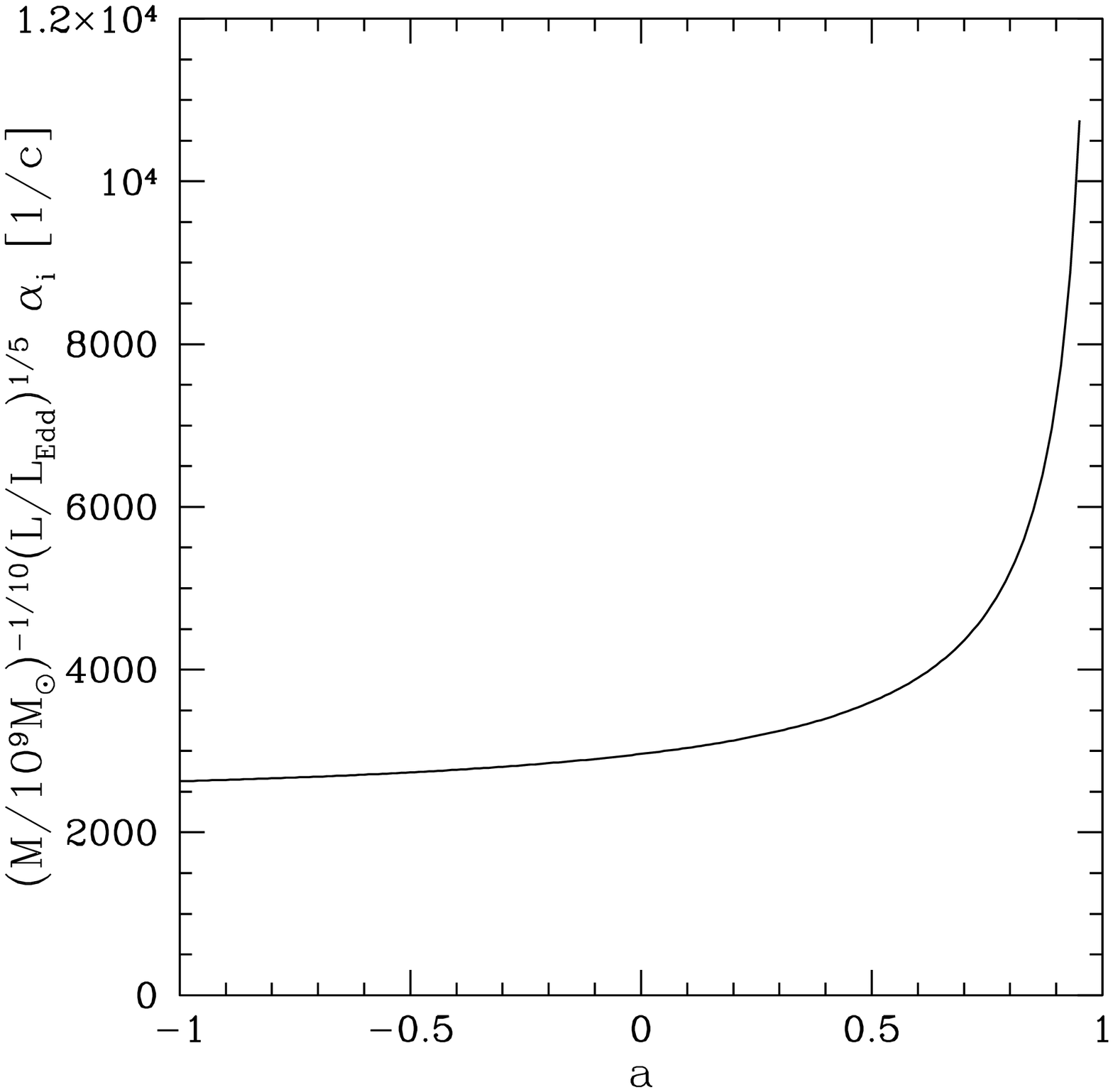}
\caption{Dependence of $\alpha_i$ on the angular momentum of the black hole,
for the $\mu = 0$ case.}
\end{figure}

\begin{figure}
\figurenum{8}
\epsscale{1}
\plotone{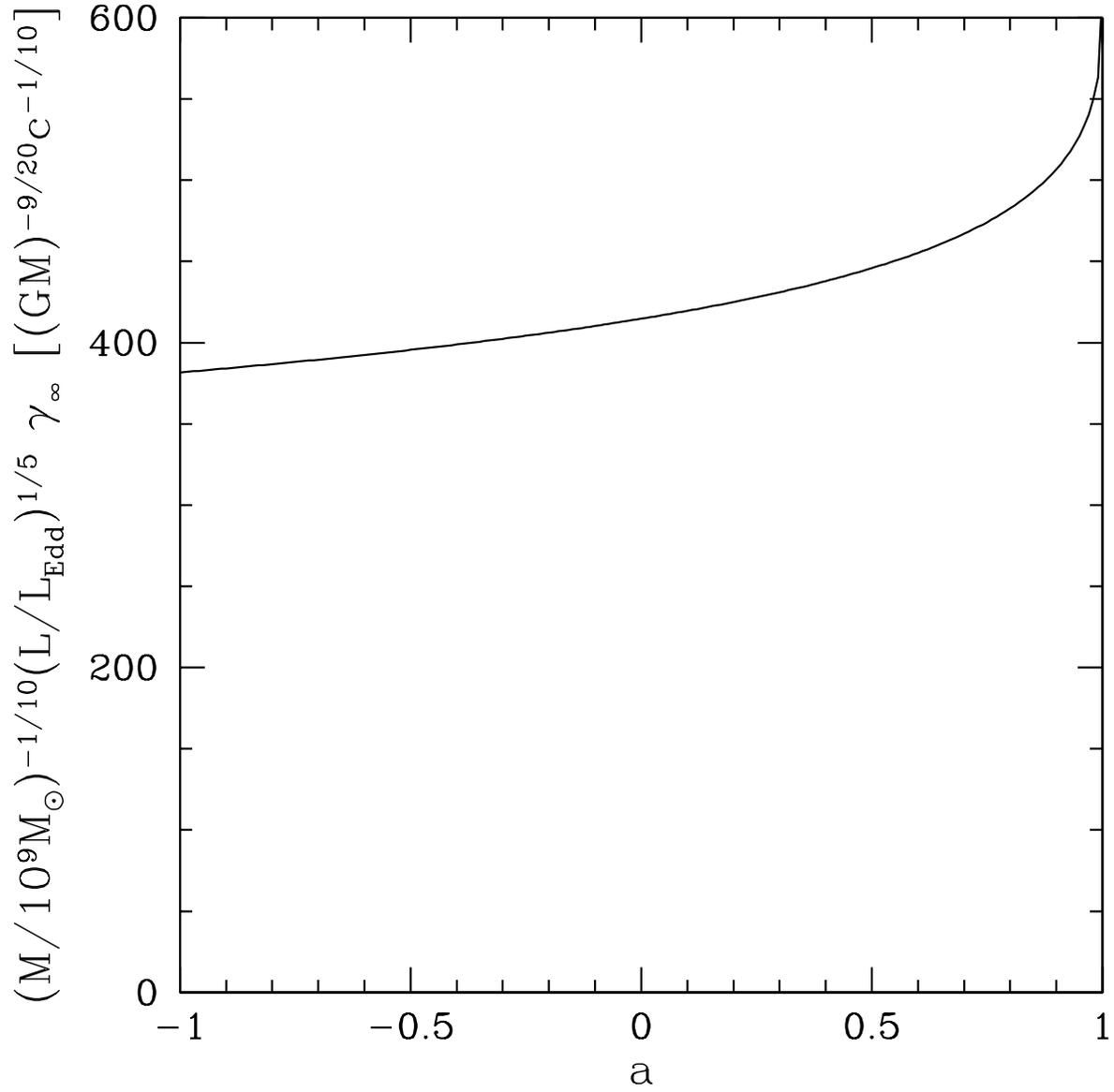}
\caption{Dependence of $\gamma_\infty$ 
on the angular momentum of the black hole.} 
\end{figure}

\begin{figure}
\figurenum{9}
\epsscale{1}
\plotone{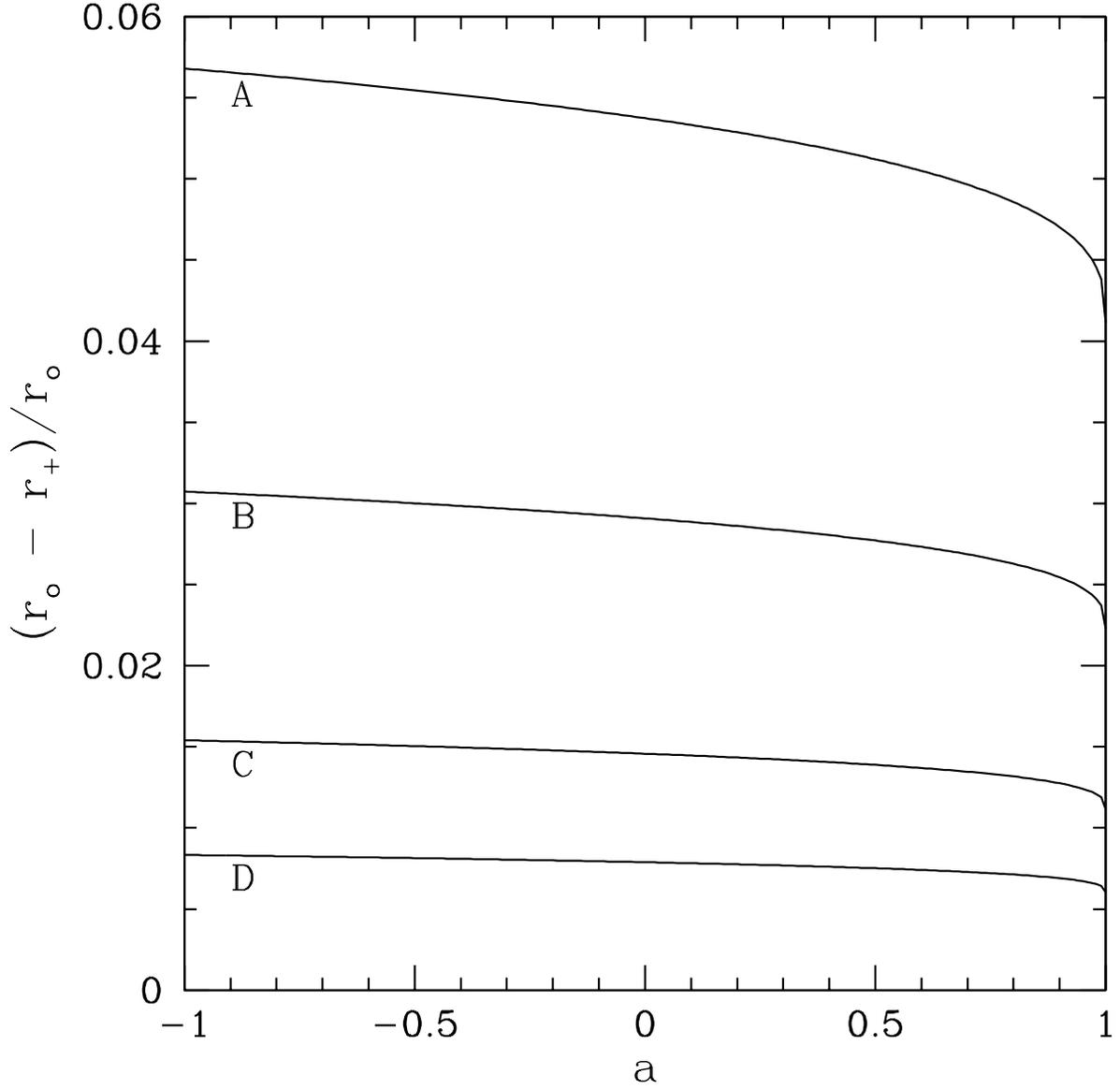}
\caption{Dependence of the fractional radial extent of fundamental outer p--modes on the angular momentum of the black hole, for
(A) $M=10 M_\sun, L = L_{Edd}, r_o = 10^4 GM/c^2$, 
(B) $M=10 M_\sun, L = 0.01 L_{Edd}, r_o = 10^4 GM/c^2$,
(C) $M=10^9 M_\sun, L = L_{Edd}, r_o = 10^3 GM/c^2$, and
(D) $M=10^9 M_\sun, L = 0.01 L_{Edd}, r_o = 10^3 GM/c^2$.}
\end{figure}

\begin{figure}
\figurenum{10}
\epsscale{1}
\plotone{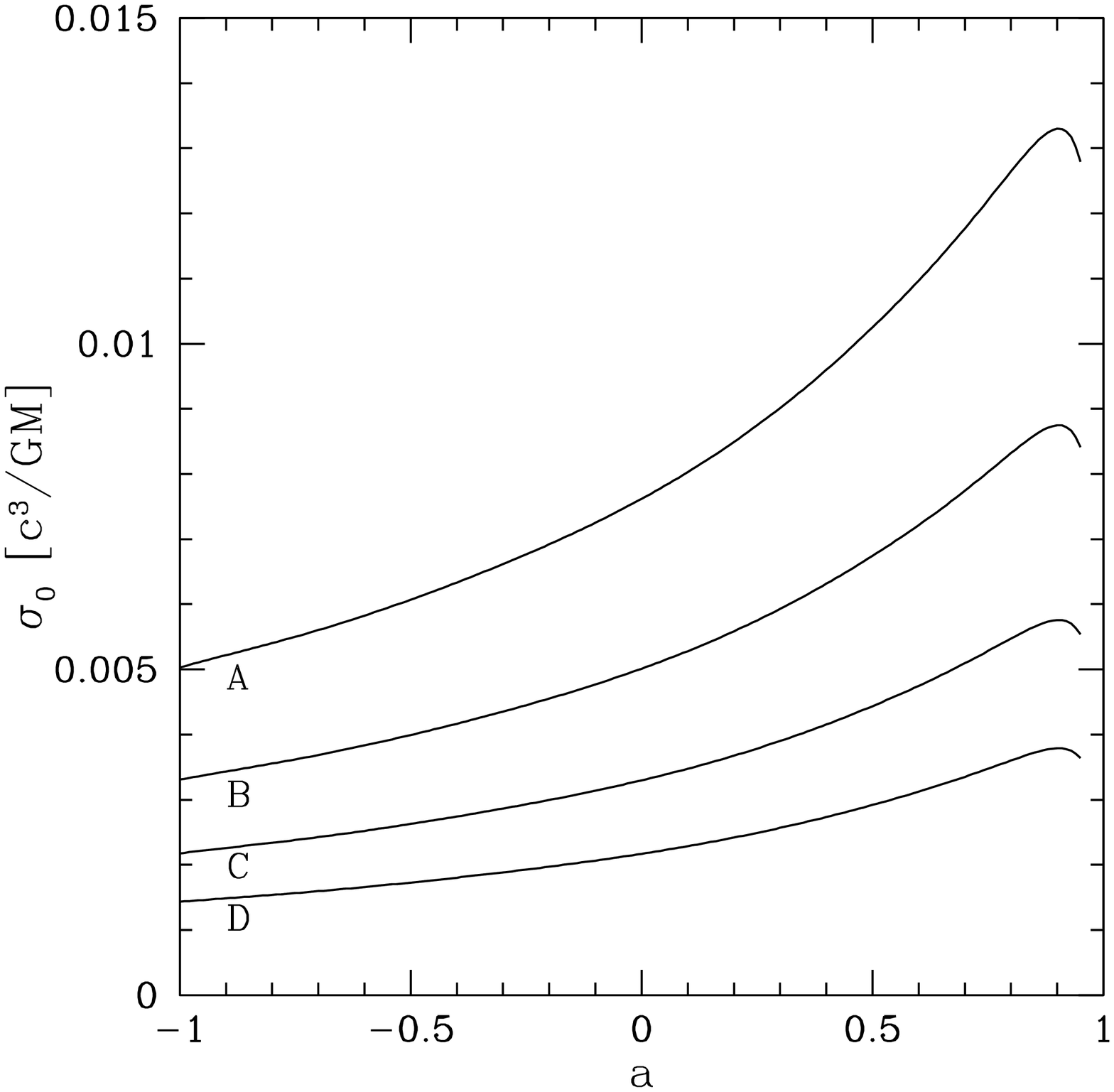}
\caption{Dependence of the eigenfrequency of $\mu = 2/5$ (no torque) fundamental inner p--modes on the angular momentum
of the black hole, for 
(A) $M=10 M_\sun, L = L_{Edd}$,
(B) $M=10 M_\sun, L = 0.01 L_{Edd}$,
(C) $M=10^9 M_\sun, L = L_{Edd}$, and
(D) $M=10^9 M_\sun, L = 0.01 L_{Edd}$.}
\end{figure}

\begin{figure}
\figurenum{11}
\epsscale{1}
\plotone{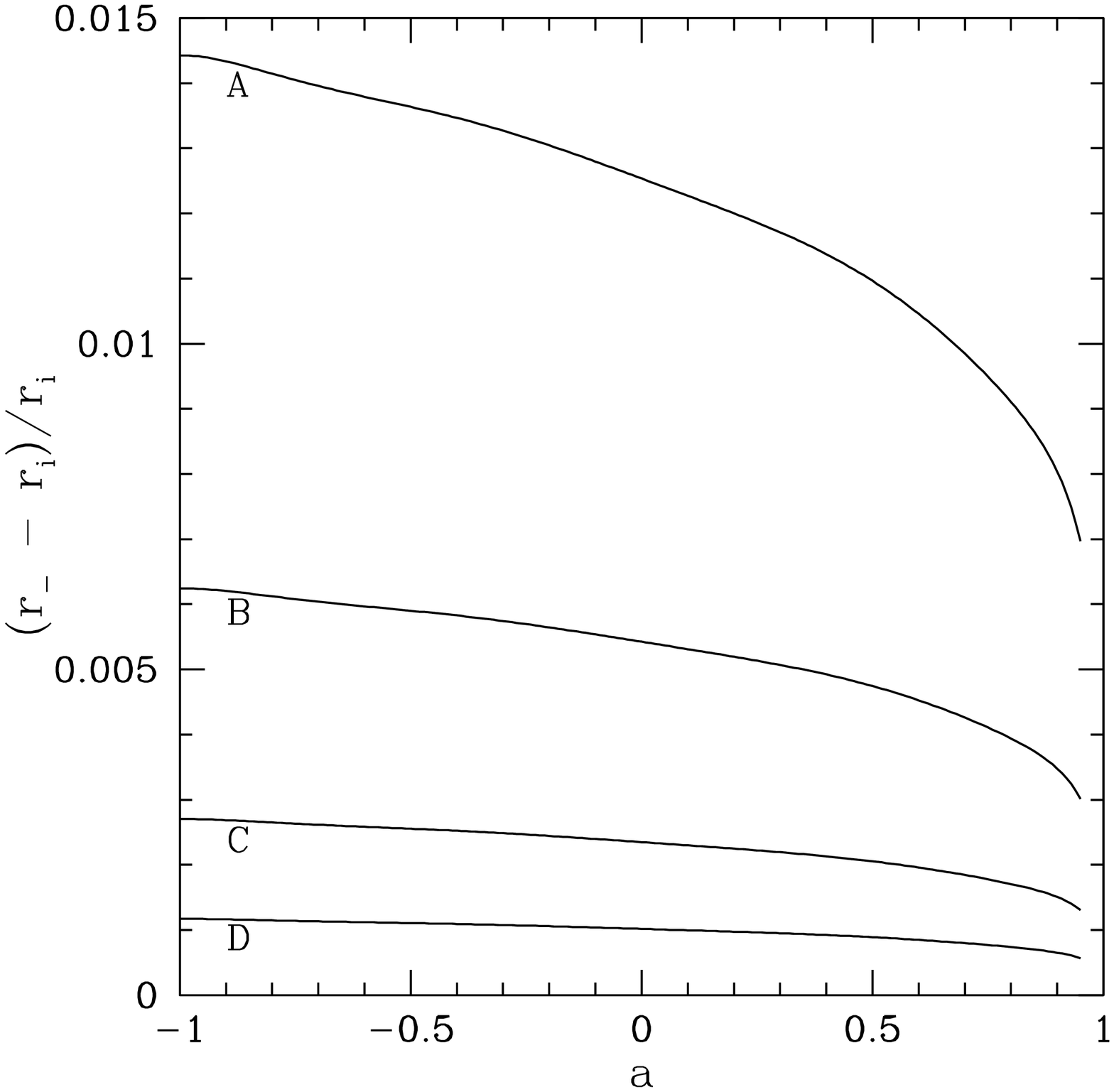}
\caption{Dependence of the fractional radial extent of $\mu = 2/5$ (no torque) fundamental inner p--modes on the angular momentum of the black hole, for 
(A) $M=10 M_\sun, L = L_{Edd}$,
(B) $M=10 M_\sun, L = 0.01 L_{Edd}$,
(C) $M=10^9 M_\sun, L = L_{Edd}$, and
(D) $M=10^9 M_\sun, L = 0.01 L_{Edd}$.}
\end{figure}

\begin{figure}
\figurenum{12}
\epsscale{1}
\plotone{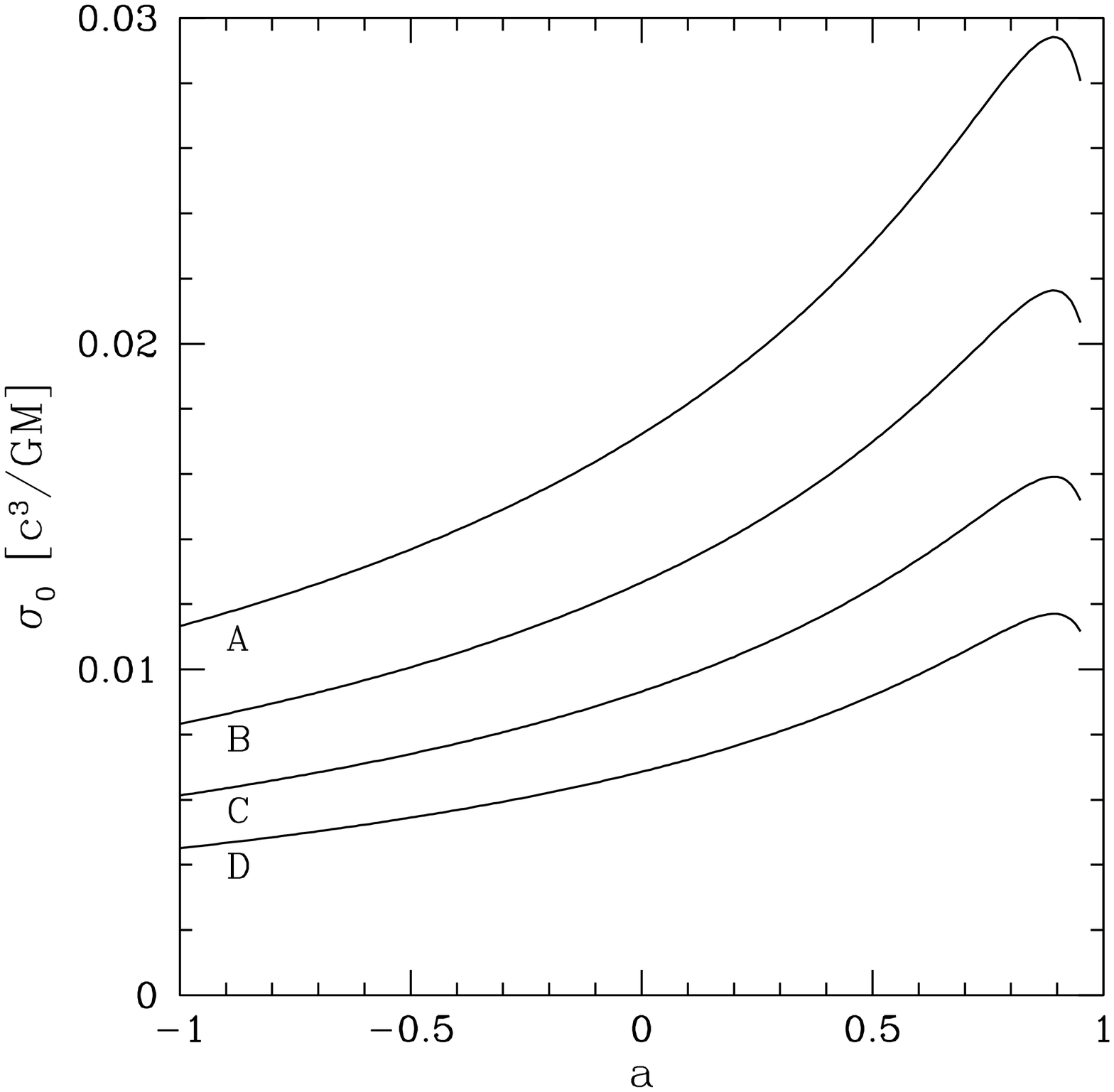}
\caption{Dependence of the eigenfrequency of $\mu = 0$, $\beta = 0.95$ fundamental inner p--modes on the angular momentum
of the black hole, for 
(A) $M=10 M_\sun, L = L_{Edd}$,
(B) $M=10 M_\sun, L = 0.01 L_{Edd}$,
(C) $M=10^9 M_\sun, L = L_{Edd}$, and
(D) $M=10^9 M_\sun, L = 0.01 L_{Edd}$.}
\end{figure}

\begin{figure}
\figurenum{13}
\epsscale{1}
\plotone{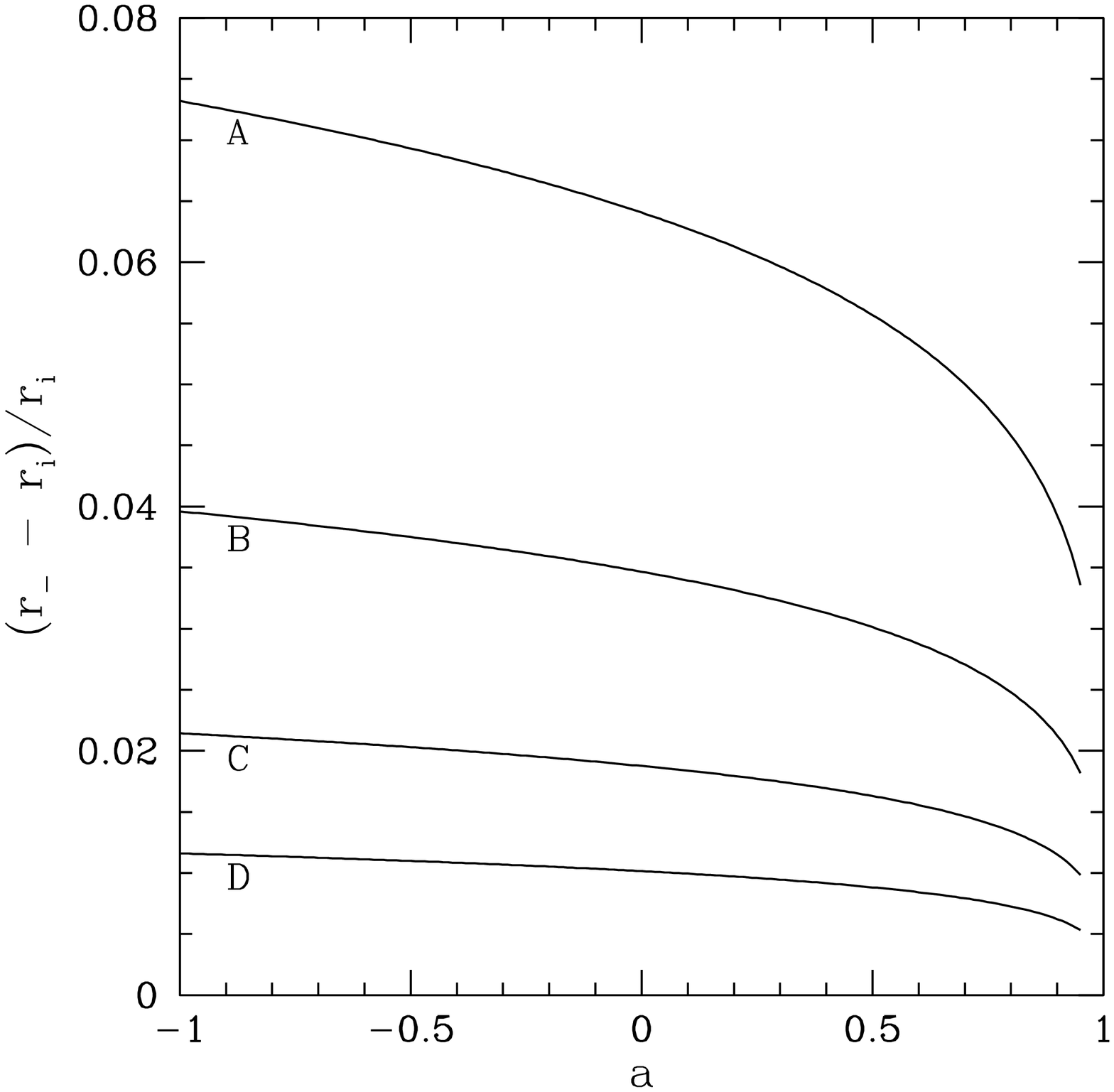}
\caption{Dependence of the fractional radial extent of $\mu = 0$, $\beta = 0.95$ fundamental inner p--modes on the angular omentum of the black hole, for 
(A) $M=10 M_\sun, L = L_{Edd}$,
(B) $M=10 M_\sun, L = 0.01 L_{Edd}$,
(C) $M=10^9 M_\sun, L = L_{Edd}$, and
(D) $M=10^9 M_\sun, L = 0.01 L_{Edd}$.}
\end{figure}

\begin{figure}
\figurenum{14}
\epsscale{1}
\plotone{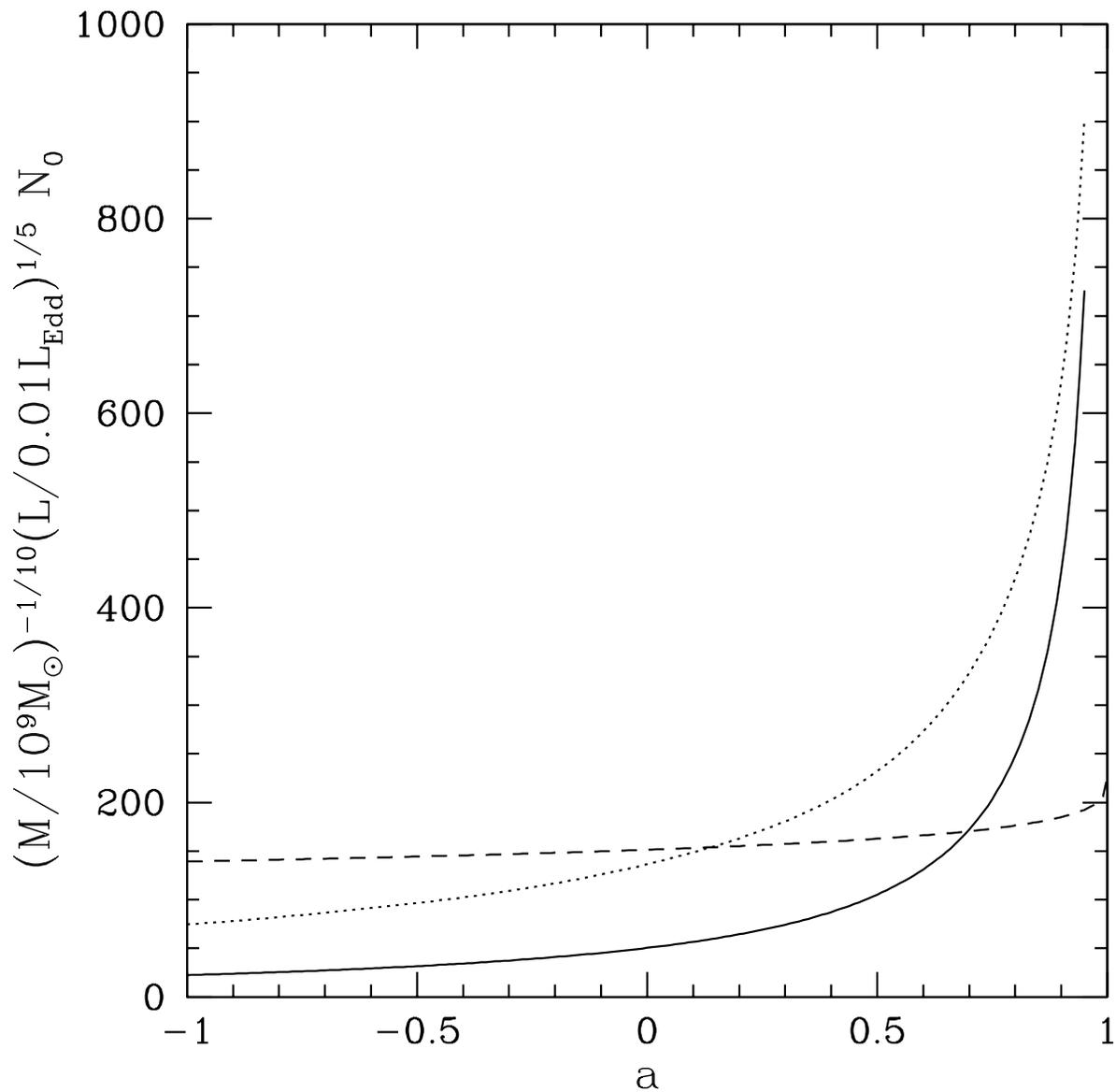}
\caption{Dependence of $N_0$
on the angular momentum of the black hole;
the number of radial modes $N$ (described by our model) satisfies $N \ll N_0$.
The three curves correspond to the
outer (broken line), $\mu = 2/5$ inner (dotted line), and 
$\mu = 0, \beta = 0.95$ inner p--modes (continuous line).
A value of $r_o = 10^4 GM/c^2$ was used for the outer modes.}

\end{figure}

\begin{figure}
\figurenum{15}
\epsscale{1}
\plotone{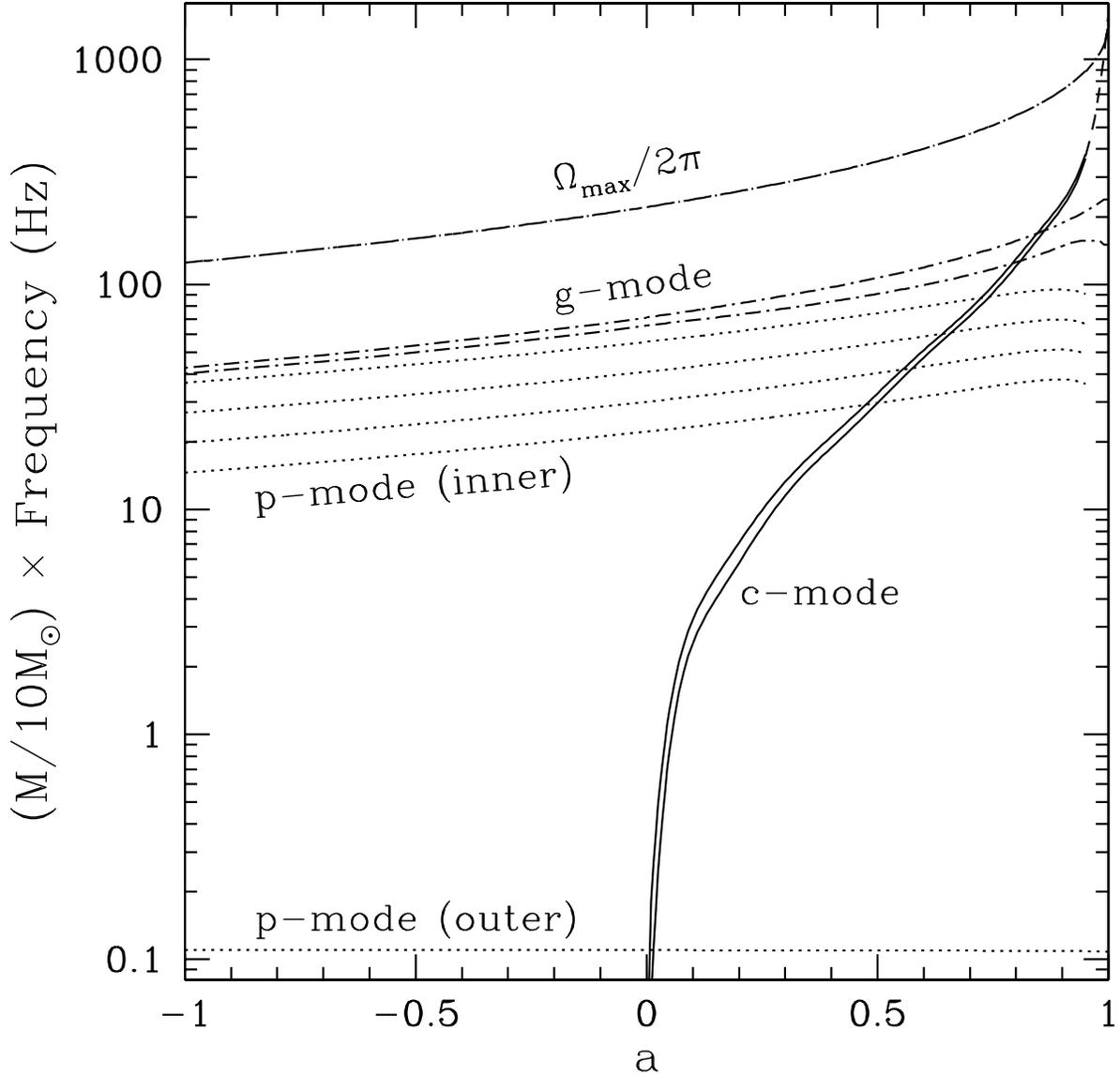}
\caption{Dependence of characteristic disk fundamental frequencies on the angular momentum of the black hole. The top one is the orbital frequency of a `blob' at the inner edge of the accretion disk. The g-- and c--mode curves bound the region corresponding to the parameters ranges $0.01\leq L/L_{Edd}\leq 1$ and $10\leq M/M_\sun\leq 10^9$.
The two uppermost inner p--mode curves bound the region corresponding to $0.01\leq L/L_{Edd}\leq 1$ and $M = 10 M_\sun$, while the bottom ones bound the region corresponding to $0.01\leq L/L_{Edd}\leq 1$ and $M = 10^9 M_\sun$.
In all cases, $\mu = 0$ with $\beta = 0.95$.
The outer p--mode curve corresponds to $L = L_{Edd}$, $M = 10 M_\sun$, and $r_o = 10^3 GM/c^2$.}
\end{figure}


\begin{thebibliography}{}

\bibitem[Hawley \& Krolik(2001)]{hk} Hawley, J.F. \& Krolik, J.H. 2001, \apj, 
548, 348
\bibitem[Honma, Matsumoto \& Kato(1992)]{hmk} Honma, F., Matsumoto, R. \& Kato, S. 1992, \pasj, 44, 529
\bibitem[Ipser \& Lindblom(1992)]{il} Ipser, J.R. \& Lindblom, L. 1992, \apj, 389, 392
\bibitem[Kato(2001)]{kato} Kato, S. 2001, \pasj, 53, 1
\bibitem[Kato et al.(1998)]{k98} Kato, S., Fukue, J., \& Mineshige, S. 1998,
Black-Hole Accretion Disks (Kyoto: Kyoto Univ. Press)
\bibitem[Kato \& Fukue(1980)]{kf} Kato, S. \& Fukue, J. 1980, \pasj, 32, 377
\bibitem[Milsom \& Taam(1997)]{mt} Milsom, J.A. \& Taam, R.E. 1997, \mnras, 286, 358
\bibitem[Novikov \& Thorne(1973)]{nt} Novikov, I.D. \& Thorne, K.S. 1973, in Black Holes, ed. C. DeWitt \& B.S. DeWitt (New York: Gordon and Breach)
\bibitem[Nowak \& Wagoner(1992)]{nw92} Nowak, M.A. \& Wagoner, R.V. 1992, \apj, 393, 697
\bibitem[Lubow \& Ogilvie(1998)]{ol} Lubow, S.H. \& Ogilvie, G.I. 1998, \apj, 504, 983
\bibitem[Ortega-Rodr\'{\i}guez, Silbergleit \& Wagoner(2001)]{osw} Ortega-Rodr\'{\i}guez, M., Silbergleit, A.S. \& Wagoner, R.V. 2001, to be submitted
\bibitem[Ortega-Rodr\'{\i}guez \& Wagoner(2000)]{ort} Ortega-Rodr\'{\i}guez, M. \& Wagoner, R.V. 2000, \apj, 537, 922
\bibitem[Page \& Thorne(1974)]{pt} Page, D.N. \& Thorne, K.S. 1974, \apj, 191, 499
\bibitem[Perez(1993)]{p} Perez, C.A. 1993, Ph.~D. thesis, Stanford Univ.
\bibitem[Perez et al.(1997)]{per} Perez, C.A., Silbergleit, A.S., Wagoner, R.V. \& Lehr, D.E. 1997, \apj, 476, 589
\bibitem[Ponomarev(1965)]{pon} Ponomarev, L.I. 1965, Sov.~Phys.-Doklady, 10, 506
\bibitem[Silbergleit, Wagoner \& Ortega-Rodr\'{\i}guez(2001)]{swo} Silbergleit, A.S., Wagoner, R.V. \& Ortega-Rodr\'{\i}guez, M. 2001, \apj, 548, 335
\bibitem[Lense \& Thirring(1918)]{lt} Lense, J. \& Thirring, H. 1918, Phys.~Z., 19, 156
\bibitem[Wagoner(1999)]{w} Wagoner, R.V. 1999, Phys. Rep., 311, 259
\bibitem[Wagoner, Silbergleit \& Ortega-Rodr\'{\i}guez(2001)]{wso} Wagoner, R.V., Silbergleit, A.S. \& Ortega-Rodr\'{\i}guez, M. 2001, \apj, 559, L25

\end{thebibliography}
\end{document}